# The angular momentum distribution and baryon content of star forming galaxies at $z$~1-3[1]

A. Burkert[1,2], N. M. Förster Schreiber[2], R. Genzel[2,3,4], P. Lang[2], L. J. Tacconi[2], E. Wisnioski[2], S. Wuyts[2,5], K. Bandara[2], A. Beifiori[1,2], R. Bender[1,2], G. Brammer[6], J. Chan[2], R. Davies[2], A. Dekel[7], M. Fabricius[1,2], M. Fossati[1,2], S. Kulkarni[2], D. Lutz[2], J. T. Mendel[1,2], I. Momcheva[8], E. J. Nelson[8], T. Naab[9], A. Renzini[10], R. Saglia[1,2], R. M. Sharples[11], A. Sternberg[12], D. Wilman[1,2] & E. Wuyts[2]

[1] *Universitäts-Sternwarte Ludwig-Maximilians-Universität (USM), Scheinerstr. 1, München, D-81679, Germany* (burkert@usm.uni-muenchen.de)

[2] *Max-Planck-Institut für extraterrestrische Physik (MPE), Giessenbachstr.1, D-85748 Garching, Germany* (genzel@mpe.mpg.de, forster@mpe.mpg.de)

[3] *Department of Physics, Le Conte Hall, University of California, Berkeley, CA 94720, USA*

[4] *Department of Astronomy, New Campbell Hall, University of California, Berkeley, CA 94720, USA*

[5] *Department of Physics, University of Bath, Claverton Down, Bath, BA2 7AY, United Kingdom*

[6] *Space Telescope Science Institute, 3700 San Martin Drive, Baltimore, MD 21218, USA*

[7] *Racah Institute of Physics, The Hebrew University, Jerusalem 91904, Israel*

[8] *Department of Astronomy, Yale University, P.O. Box 208101, New Haven, CT 06520-810, USA*

[9] *Max-Planck Institute for Astrophysics, Karl Schwarzschildstrasse 1, D-85748 Garching, Germany*

[10] *INAF-Osservatorio Astronomico di Padova, Vicolo dell'Osservatorio 5, Padova, I-35122, IT*

[11] *Department of Physics, Durham University, Science Laboratories, South Road Durham DH1 3LE, UK*

[12] *School of Physics and Astronomy, Tel Aviv University, Tel Aviv 69978, Israel*

---

[1] based on observations obtained at the Very Large Telescope of the European Southern Observatory, Paranal, Chile (ESO Programme IDs 075.A-0466, 076.A-0527, 079.A-0341, 080.A-0330, 080.A-0339, 080.A-0635, 081.B-0568, 081.A-0672, 082.A-0396, 183.A-0781, 087.A-0081, 088.A-0202, 088.A-0209, 091.A-0126, 092.A-0091, 093.A-0079, 094.A-0217, 095.A-0047, 096.A-0025).

# Abstract

We analyze the angular momenta of massive star forming galaxies (SFGs) at the peak of the cosmic star formation epoch ($z \sim 0.8 - 2.6$). Our sample of ~360 $\log(M_*/M_\odot) \sim 9.3 - 11.8$ SFGs is mainly based on the KMOS$^{3D}$ and SINS/zC-SINF surveys of Hα kinematics, and collectively provides a representative subset of the massive star forming population. The inferred halo scale angular momentum distribution is broadly consistent with that theoretically predicted for their dark matter halos, in terms of mean spin parameter $\langle\lambda\rangle \sim 0.037$ and its dispersion ($\sigma_{\log \lambda} \sim 0.2$). Spin parameters correlate with the disk radial scale, and with their stellar surface density, but do not depend significantly on halo mass, stellar mass, or redshift. Our data thus support the long-standing assumption that on average, even at high redshifts, the specific angular momentum of disk galaxies reflects that of their dark matter halos ($j_d = j_{DM}$). The lack of correlation between $\lambda \times (j_d/j_{DM})$ and the nuclear stellar density $\Sigma_*(1\text{kpc})$ favors a scenario where disk-internal angular momentum redistribution leads to 'compaction' inside massive high-redshift disks. For our sample, the inferred average stellar-to-dark matter mass ratio is ~2%, consistent with abundance matching results. Including the molecular gas, the total baryonic disk-to-dark matter mass ratio is ~5% for halos near $10^{12}$ $M_\odot$, which corresponds to 31% of the cosmologically available baryons, implying that high-redshift disks are strongly baryon dominated.

# 1. Introduction

In the cold dark matter paradigm, baryonic disk galaxies form at the centers of dark matter halos (e.g., Fall & Efstathiou 1980; Fall 1983; see Mo et al. 2010 for a review). Defining the halo radius as the region within which the virialized dark matter particles have on average 200 times the mean mass density of the Universe, the halo's virial velocity, $v_{virial}$, its mass $M_{DM}$, and its virial radius $R_{virial}$ are given in the spherical collapse model (for a flat $\Lambda$CDM Universe) by the following well known relations (Peebles 1969; Gunn & Gott 1972; Bertschinger 1985; Mo et al. 1998)

$$R_{virial} = \frac{v_{virial}}{10H(z)} \text{ and}$$

$$M_{DM} = \frac{v_{virial}^2 R_{virial}}{G} = \frac{v_{virial}^3}{10GH(z)}, \text{ where}$$

$$H(z) = H_0(\Omega_{\Lambda,0} + \Omega_{m,0} \times (1+z)^3)^{1/2} \qquad (1).$$

Here $G$ is the gravitational constant, $H(z)$ and $H_0$ are the Hubble constants at $z$ and $z=0$, and $\Omega_{\Lambda,0}$ and $\Omega_{m,0}$ are the energy densities of $\Lambda$ and total matter at $z = 0$, relative to the closure density. Tidal torque theory (Hoyle 1951; Peebles 1969; White 1984) suggests that within the virial radius, the centrifugal support of baryons and dark matter (labeled 'DM' from hereon) is small and given by the spin parameter,

$$\lambda = \frac{\omega_{virial}}{\omega_{virial,cs}} = \varepsilon \frac{J_{DM}/M_{DM}}{R_{virial} \times v_{virial}} \sim \frac{J_{DM} \times E_{DM}^{1/2}}{GM_{DM}^{5/2}} \qquad (2),$$

where $\omega=v_{rot}/R$ is the angular speed ($v_{rot}$ is the rotational/tangential velocity) at $R$, and 'virial' and 'cs' stand for 'within the virial radius' and 'centrifugal support' ($\omega_{rot,cs}=(GM/R^3)^{1/2}$). The constant $\varepsilon$ is $\sim \sqrt{2}$, $J$ and $j$ are the total and specific ($j=J/M$) angular momenta, and $E\sim GM^2/R$ is the absolute value of the total gravitational energy. Building on earlier work by Peebles (1969) and Barnes & Efstathiou (1987), simulations have shown that tidal torques generate a universal, near-lognormal distribution function of halo spin parameters, with $<\lambda>=0.035$-$0.05$ and a dispersion of $\pm 0.2$ in the log (Bullock et al. 2001a; Hetznecker & Burkert 2006; Bett et al. 2007; Maccio et al. 2007).

If the baryons are dynamically cold, or they can cool after shock heating at $R_{virial}$, they fall inwards and form a centrifugally supported disk of (exponential) radial scale length $R_d$, given by (e.g., Mo et al. 1998; see also Fall 1983, their Equation 4)

$$R_d = \frac{1}{\sqrt{2}}\left(\frac{f_{Jd}}{m_d}\right) \times \lambda \times R_{virial} = \frac{1}{\sqrt{2}}\left(\frac{j_d}{j_{DM}}\right) \times \lambda \times R_{virial} \qquad (3).$$

Here $m_d = M_d/M_{DM}$ is the ratio of the baryonic disk mass to that of the dark matter halo and $f_{Jd}$ is the fraction of the total dark halo angular momentum in the disk, $J_d = f_{Jd} J_{DM}$.

In the literature it has generally been assumed that the specific angular momentum of the baryons and the dark matter is the same, such that $j_d = j_{DM}$ (e.g., Dutton & van den Bosch 2012). Indeed, models adopting $j_d = j_{DM}$ have been very successful in explaining the scaling relations of low-redshift disk galaxies (e.g. Fall 1983; Mo et al. 1998; Dutton & van den Bosch 2012; Romanowsky & Fall 2012; Fall & Romanowsy 2013). The situation is however different for passive spheroids. Fall & Romanowsky (2013) found $j_d/j_{DM} \sim 0.8$ for late-type, star forming disks, but only $\sim 0.1$ for early-type passive spheroids, with Sa and S0 galaxies in between these two extremes. Early numerical simulations of cosmological disk galaxy formation suffered from catastrophic angular momentum loss, leading to disk galaxies with scale lengths that were an order of magnitude smaller than observed (Navarro & Benz 1991; Navarro & White 1994; Navarro & Steinmetz 1997). More recent simulations using improved numerical schemes and including stellar feedback however confirmed the assumption $j_d = j_{DM}$ (e.g. Übler et al. 2014; Danovich et al. 2015; Teklu et al. 2015). We note however that this result is not at all trivial. The infalling baryons can both lose and gain angular momentum between the virial and disk scale. In addition, the baryon fraction of galaxies, including disk galaxies, is much smaller than the cosmic baryon fraction, indicating that substantial amounts of gas either never entered the galactic plane or were blown out afterwards. In this case, the specific angular momentum of the gas that is retained in the disks could be very different compared to the specific angular momentum of the gas entering the virial radius.

Most studies so far concentrated on galaxies in the low-redshift Universe. Recent high-resolution simulations of high-redshift disk galaxy formation by Danovich et al. (2015) found that the gas entering the virial radius in cold streams has $\sim 3 \lambda_{DM}$. Subsequent angular momentum redistribution and loss by torques and feedback-driven outflows however leads to disk spins that are similar to the halo spins. Clearly, given this complexity, it is of great interest to empirically study the baryonic angular momentum distributions of galaxies as a function of cosmic epoch.

Another important physical parameter of galaxy formation is the relative fraction $m_d$ of baryonic-to-dark matter mass in the half-light regions $R_{1/2}$ of $z \sim 0$ galaxies, which depends on type and mass. Massive early-type spheroidal systems and massive disks, including the Milky Way, are baryon dominated within $\sim 1.2$ $R_{1/2}$ (called 'maximal disks' if $M_{DM}/M_{baryon} < 0.3$ within that radius; Courteau & Dutton 2015; Barnabè et al. 2012; Dutton et al. 2013; Cappellari et al. 2013; Bovy & Rix 2013). In contrast, the dark matter fraction is significant and becomes even dominant for dwarf spheroidal galaxies and lower mass disks (Martinsson et al. 2013a,b). In the outer regions (on scales of 10-30 kpc) $z = 0$ disks are dark matter dominated, as demonstrated by their flat rotation curves (e.g., Sofue & Rubin 2001; Courteau & Dutton 2015).

At high redshift little is known *empirically* so far about the baryonic angular momentum distribution (see Förster Schreiber et al. 2006 for a first attempt). Lookback studies have shown that most 'normal', massive star forming galaxies (selected from rest-frame UV/optical imaging surveys) from $z \sim 0$ to $z \sim 3$ are located on or near a star formation 'main sequence' in the stellar mass ($M_*$) versus star formation rate (SFR) plane. Its slope is approximately independent of redshift and slightly sublinear (SFR $\sim M_*^{0.7-1}$), but its amplitude strongly increases with redshift to $z \sim 2.5$ (such that the specific star formation rate sSFR = SFR/$M_*$ $\sim (1+z)^{2.5-3}$; Daddi et al. 2007; Noeske et al. 2007; Schiminovich et al. 2007; Rodighiero et al. 2010, 2011; Whitaker et al. 2012, 2014; Speagle et al. 2014). The location of galaxies in the stellar mass-specific star formation rate plane correlates with their internal structure. Out to at least $z \sim 2.5$, typical star forming galaxies (SFGs) on the 'main sequence' are well approximated by exponential light and mass profiles with Sérsic index $n_S \sim 1$ while passive galaxies below the main sequence, outlier starbursts well above the main sequence, as well as the most massive (log($M_*/M_\odot$) >11) main sequence SFGs tend to exhibit cuspier profiles with Sérsic indices $n_S > 2$ (e.g., Wuyts et al. 2011b, 2012; Bell et al. 2012; Lang et al. 2014; Bruce et al. 2014a,b; Nelson et al. 2015; Whitaker et al. 2015).

The ionized gas kinematics of these SFGs are broadly consistent with these structural properties (e.g., Genzel et al. 2006, 2008; Förster Schreiber et al. 2009;

Wright et al. 2009; Law et al. 2009; Jones et al. 2010; Épinat et al. 2009, 2012; Wisnioski et al. 2015; see Glazebrook 2013 for a more complete review). The majority (> 70%) of massive ($\log(M_*/M_\odot)$ >10) main-sequence SFGs at $z \sim 2.5$ are rotationally supported disks (e.g., Newman et al. 2013; Wisnioski et al. 2015), albeit with large velocity dispersions and often clumpy and irregular rest-frame UV/optical morphologies (Cowie et al. 1995, 1997; van den Bergh et al. 1996; Elmegreen et al. 2004, 2009; Elmegreen 2009; Förster Schreiber et al. 2011b; Wuyts et al. 2012).

In this paper we want to take the next step and explore the angular momentum distribution and baryon to dark matter fractions in $z \sim 0.8 - 2.6$ star forming galaxies, at the peak of cosmic star formation activity, by taking advantage of the recent growth in sample sizes and coverage of the $M_*$ - SFR plane with H$\alpha$ kinematics integral field unit (IFU) data sets. This progress has started in the last few years, for instance with the SINS/zC-SINF (e.g., Förster Schreiber et al. 2006, 2009; Mancini et al. 2011; N. M. Förster Schreiber et al. 2016, in preparation), MASSIV (e.g., Épinat et al. 2009, 2012; Contini et al. 2012) and HiZELS (e.g., Swinbank et al. 2012) surveys with SINFONI on the VLT, as well as with surveys with OSIRIS on the Keck telescope (e.g., Law et al. 2009; Wright et al. 2009; WiggleZ, Wisnioski et al. 2011, 2012). Most importantly we have recently started the KMOS$^{3D}$ survey (e.g., Wisnioski et al. 2015) with the multiplexed near-infrared IFU spectrometer KMOS on the VLT (Sharples et al. 2008, 2012), which will deliver IFU data for at least $\sim 600$ $z \sim 0.6$-2.7 SFGs (see also, e.g., Sobral et al. 2013; Stott et al. 2014, 2016; Mendel et al. 2015; Magdis et al. 2016; Harrison et al. 2016, for other examples of KMOS surveys of distant galaxies). The combined data of these surveys currently provide a sample of over 1000 galaxies, with a good coverage of massive ($\log(M_*/M_\odot)$ >10) star forming galaxies in the $z \sim 0.8 - 2.6$ redshift range.

Throughout the paper, we adopt a flat $\Lambda$CDM cosmology with $\Omega_{m,0}$=0.27, $\Omega_{b,0}$=0.046 and $H_0$=70 km/s/Mpc (Komatsu et al. 2011), and a Chabrier (2003) initial stellar mass function (IMF).

## 2. Observations and Analysis

### 2.1. Galaxy Sample

We base this study on IFU observations of the Hα kinematics and distribution in a large initial sample of 433 $z = 0.76 - 2.6$ massive, star forming disk galaxies. The data for these galaxies come from different IFU surveys, either ongoing or in the literature, with the 2-year sample of the KMOS$^{3D}$ survey (Wisnioski et al. 2015) constituting the strong majority (~3/4: 316 of the 433 galaxies). Since the subject of this study is an analysis of angular momenta, we first eliminated from the initial sample all major mergers (23 galaxies), all dispersion-dominated galaxies ($v_{rot}/\sigma_0 < 1.5$, 31 galaxies; see below), and 20 galaxies without well-defined kinematics and/or with very large beam smearing corrections (see below). This leaves us with a disk sample of 359 SFGs, which we will denote henceforth as the '*full*' sample. We also created a second, still more restricted '***best***' sample of the 233 highest quality, well resolved rotating disks, by retaining only $v_{rot}/\sigma_0 \geq 2$ SFGs. We also eliminated minor mergers and SFGs with obviously perturbed morphologies/kinematics, as well as galaxies with an offset between the morphological major axis (continuum or Hα) and the kinematic major axis of greater than 40°. Finally we culled all insufficiently resolved disks with a half-light/mass radius $R_{1/2} < 2$ kpc and a ratio of $R_{1/2}$ to the HWHM beam size $R_{1/2,beam}$ less than unity, and SFGs with still significant beam smearing corrections. SFGs in this 'best' sample exclusively come from SINS/zC-SINF and KMOS$^{3D}$ and were all analyzed in a consistent manner.

Figure 1 shows that our final IFU sample yields a good representation of the mass-selected main-sequence SFG population at $\log(M_*/M_\odot)>10.1$ and sSFR/sSFR(ms,$z$) > 0.1 in the range $0.8 \leq z \leq 2.6$, as drawn from the 3D-HST survey catalogs (Brammer et al. 2012; Skelton et al. 2014; Momcheva et al. 2015) in the CANDELS extragalactic survey fields (Grogin et al. 2011; Koekemoer et al. 2011) – hereafter 'reference' galaxy sample or population. Our kinematic sample by design is heavily incomplete for lower mass SFGs ($\log(M_*/M_\odot) <10$) and does not cover the passive population. The inhomogeneous redshift coverage in the upper left panel of Figure 1, with a significant lack of SFGs at $1.2 < z < 2$, reflects the intervals where Hα is shifted between the *J*- and *H*-, and *H*- and *K*-band atmospheric transmission windows, and the emphasis on $z \sim 0.76 - 1.1$ and $z \sim 2-2.6$ slices in the first two years of the KMOS$^{3D}$ survey (Wisnioski et al. 2015).

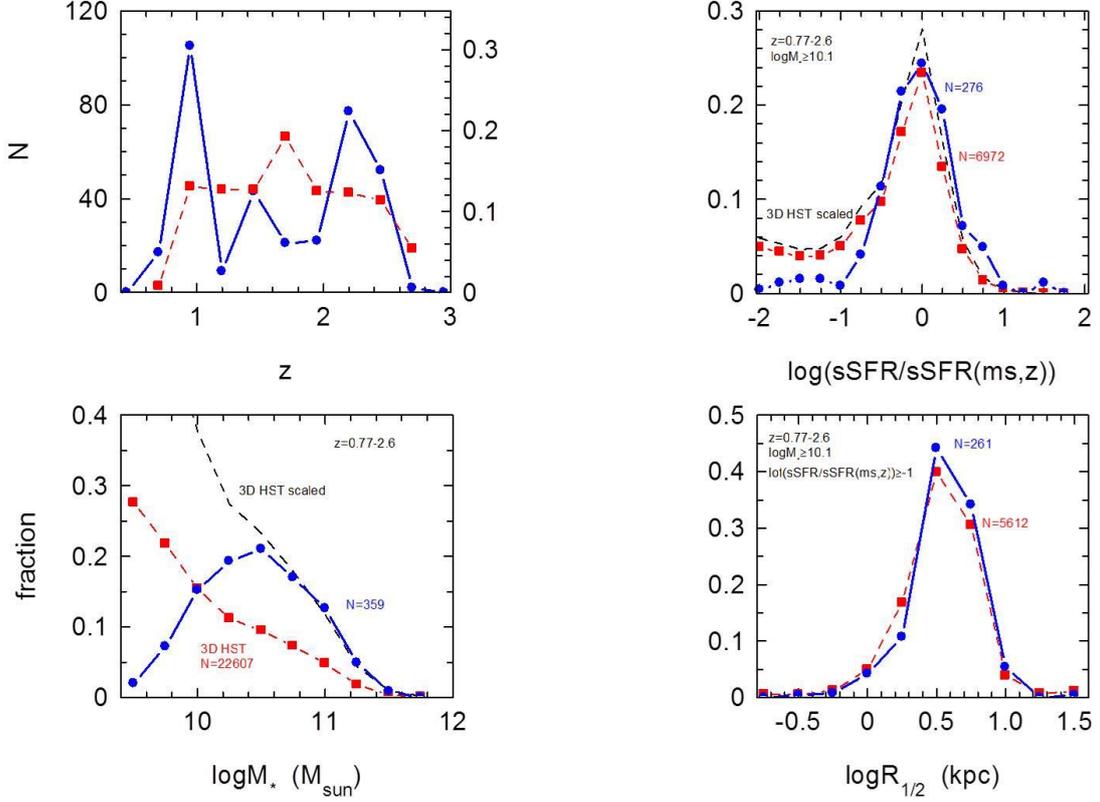

**Figure 1.** Redshift (top left), stellar mass (bottom left), specific star formation rate (top right, relative to the main sequence relation at a given $z$), and disk half-light radius (bottom right) distributions of $0.7 < z < 2.7$ galaxies from the 3D-HST source catalogs (red squares and dashed lines; Brammer et al. 2012, Skelton et al. 2014; Momcheva et al. 2015) in the CANDELS extragalactic survey fields (Grogin et al. 2011; Koekemoer et al. 2011), and the 'full' Hα IFU sample assembled for this paper (filled blue circles and solid lines). The left panels refer to all 359 SFGs in our sample. The upper right panel compares all galaxies with $\log(M_*/M_\odot) > 10.1$ (above which our sample is a good representation of the 3D-HST sample), leaving 276 SFGs. The bottom right panel compares the size distributions of the 3D-HST and Hα samples in that part of the stellar mass–star formation rate plane where our sample is representative of the 3D-HST reference sample ($\log(M_*/M_\odot) > 10.1$, sSFR/sSFR(ms,$z$) $> 0.1$, 261 SFGs). The black dashed line in the bottom left panel shows the same 3D-HST distribution as plotted in red but scaled so as to match that of our Hα IFU sample (shown in blue) above $\log(M_*/M_\odot) = 10.5$; similarly, the black dashed line in the top right panel shows the 3D-HST distribution now restricted to $\log(M_*/M_\odot) > 10.1$ and scaled to match our IFU sample above sSFR/sSFR(ms,$z$) = 0.1.

Perhaps most importantly for the results in this paper, the bottom right panel of Figure 1 demonstrates that the ***distribution of galaxy half-light radii is indistinguishable from that of the underlying reference galaxy population***, when selected according to the same redshift, stellar mass, and specific star formation rate cuts. Our Hα kinematic galaxy sample thus is ***unbiased*** in terms of size and placement relative to the main sequence line.

Appendix A1 gives more details on the source selection, and Appendix A2 summarizes how the key global, structural, and kinematic properties of the galaxies were derived (including stellar and gas masses, SFRs, half-light radii $R_{1/2}$, rotation velocities $v_{rot}$, and velocity dispersions $\sigma_0$). The IFU observations, and the Hα kinematic maps, profiles, and basic measurements have been presented and discussed in detail in the main reference papers describing the surveys considered for our study; we refer the interested reader to these papers for extensive examples of the data (Wright et al. 2007, 2011; van Starkenburg et al. 2008; Förster Schreiber et al. 2009; Law et al. 2009; Épinat et al. 2009, 2012; Mancini et al. 2011; Contini et al. 2012; Swinbank et al. 2012; Genzel et al. 2013, 2014; Wisnioski et al. 2015; S. Wuyts et al. 2016; N. M. Förster Schreiber et al. 2016, in preparation).

## *2.2. Estimating the Halo Masses and Halo λ Parameters*

With the disk parameters in hand ($R_{1/2}$, $v_{rot}= v_{rot}(R\sim R_{1/2})$, $\sigma_0=\sigma (R\sim 2\ R_{1/2})$, $M_*$, $M_{gas}=M_{molgas}{}^2$, $M_{baryon}(R_{1/2})=(M_*+M_{gas})(R_{1/2})$ ), the next step is to estimate the halo masses, spin parameters, and specific disk angular momenta for the individual galaxies. Following Mo et al. (1998) we used four independent methods to reach this goal:

1. our primary approach is to determine the dark matter halo mass and angular momentum parameter, as well as the baryonic to dark matter mass ratio, from fitting an exponential disk embedded in a Navarro-Frenk-White halo (NFW;

---

[2] As explained in Appendix A2, our $M_{gas}$ estimates are derived from the scaling relations between $M_{gas}$, $M_*$, SFR, and $z$ for main sequence SFGs presented by Genzel et al. (2015), assuming that at $z\sim$1-3 the molecular component dominates and the atomic fraction can be neglected. As such, the $M_{gas}$ masses estimates may be lower limits.

Navarro et al. 1997), extracting the optimum parameters from a Monte-Carlo (MC) search of the parameter space;

2. we also estimated the halo angular momentum parameter for a dark matter dominated, isothermal halo;

3. we determined the dark matter angular momentum parameter from fitting an exponential disk embedded in an NFW halo, this time adopting a constant baryonic disk mass to dark matter halo mass ratio, and

4. as a variant, we determined the dark matter mass and angular momentum parameter from fitting an exponential disk embedded in an NFW halo, by inverting the stellar mass to dark matter mass ratio as a function of halo mass from the abundance matching work of Moster et al. (2013) and Behroozi et al. (2013a).

We discuss in Appendix B the details of our methods, including correction of the observed rotation velocities and disk sizes for asymmetric drift, and deviations from pure exponential surface density distributions.

The four methods yield independent estimates of the mass, radius, and angular momentum parameter of the dark matter halo from the mass, size, and kinematics of the central baryonic disk. All assume implicitly that the specific angular momentum of the baryons on the scale of the dark halo is the same as that of the dark matter component. More importantly and precisely, all methods deliver an estimate of the product

$\lambda \times (j_d/j_{DM}) = \lambda_{DM}(R \sim R_{virial}) \times (j_{baryon}/j_{DM})(R \sim R_{virial}) \times (j_{baryon}(R \sim R_{1/2,disk})/j_{baryon}(R \sim R_{virial}))$.

Hence, our results depend on the angular momentum distribution on the halo scale (of both baryons and dark matter), as well as on any re-distribution of angular momentum between different baryonic components (inner and outer disk, outflow, bulge, etc.). The former is a measure of the total angular momentum state of the halo ('nature'), while the latter depends on intra-halo baryonic processes ('nurture').

# 3. Results and Discussion

## *3.1. The Specific Angular Momenta of z=1-2 Star Forming Disk Galaxies*

### *3.1.1. Correlation of Angular Momenta with Stellar Mass*

In this section we begin with an empirical investigation of the observed **baryonic** angular momentum distributions of our disks. Following Romanowsky & Fall (2012) ***we assume for simplicity that all disks in our sample have the same dark matter angular momentum parameter,*** $\lambda_{0.035}=\lambda/0.035$, and we then use equations (1) and (3) to express the specific angular momentum of the stellar/baryonic disk, $j_d$, as

$$\left(\frac{j_d}{\text{km/s kpc}}\right) = 1177 \times \left(\frac{j_d}{j_{DM}}\right) \times f_*^{-2/3} \times \left(\frac{H(z)}{H_0}\right)^{-1/3} \times \lambda_{0.035} \times \left(\frac{M_*}{10^{11} M_\odot}\right)^{2/3} \quad (4).$$

Here $f_*=M_*/(0.17 \times M_{DM})$ is the fraction of the cosmologically available baryons that are tied up in the stellar disk. The assumption of constant $\lambda$ for all SFGs cannot be correct for each SFG. However, it is plausibly correct on average since the halo spin parameters are expected to follow a lognormal distribution about the mean (see Section 1), such that the average trend of the data as a function of stellar mass can then be compared to theory and other observations. The assumption of constant $\lambda$ in equation (4) should just lead to a scatter in the data, but no trends with parameters such as $M_*$, $z$, etc.

Equation (4) ties an easily observable quantity ($j_d$) to the product of the another easily observable quantity, $M_*^{2/3}$, and the ratio $j_d/j_{DM}$. For a disk of constant rotation velocity $v_{rot}$ and of effective radius $R_{1/2}$ the specific angular momentum is

$$j_d = k_d \times v_{rot} \times R_{1/2} \quad (5),$$

with the constant $k_d = 1.19$ for a thin exponential disk (Sérsic index $n_S=1$). The disks of our sample are thick, dispersion truncated, their rotation velocities are not constant, and their Sérsic indices $n_S \gtrsim 1$ (see Appendices A and B), in which case $k_d$ becomes a function of $n_S$ and $v_{rot}/\sigma_0$. This case is discussed in Appendix B.7 and we used the fitting function given in equation (B11). Since the correction function in equation (B11) uses Sérsic indices derived from the optical continuum stellar light as well as $v_{rot}/\sigma_0$ values inferred from the ionized gas kinematics, and equation (5) also mixes information obtained from stars ($R_{1/2}$) and gas ($v_{rot}$), in what follows we assume $j_d = j_*= j_{gas}$. The 3D-HST work of Nelson et al. (2015) shows that this assumption is to first order correct when comparing the disk sizes of gas and stars, but that the ionized gas disks tend to be somewhat larger than those of the stars, most notably at the highest stellar masses, where stellar bulges become prominent (($R_{1/2,H\alpha}/R_{1/2,*}$) = $1.1\times(M_*/10^{10}M_\odot)^{0.05}$).

In the left panel of Figure 2 we plot the specific angular momenta of our SFGs as a function of stellar mass, adopting equations (4) and (5), and $\lambda$ = 0.035 for all galaxies. The data follow the theoretically expected $M_*^{2/3}$ dependence. There is no significant difference between the 'full' and 'best' samples, other than that the 'best' sample by design (section 2.1) lacks a number of low mass, small SFGs with low rotational support, including dispersion-dominated galaxies. The dispersion of the data around the trend-line is ±0.031 dex for the 'best' and ±0.035 dex for the 'full' sample.

In right panel of Figure 2 we emphasize as filled red circles SFGs in the upper 25% percentile of central stellar surface density $\Sigma_*(R \leq 1$ kpc$) > 10^{9.7}$ $M_\odot$kpc$^{-2}$ (equivalent to $n_S > 3$). The observed specific angular momenta of SFGs, as traced by the ionized gas distribution at/near $R_{1/2}$, do not strongly depend on central stellar surface density, or Sérsic index. The densest, cuspiest galaxies are more common among the more massive ($\log(M_*/M_\odot) > 10.6$) SFGs, consistent with massive bulges being present (e.g., Lang et al. 2014) but the specific angular momentum of these systems appears uncorrelated with their central stellar properties. This suggests that the formation/presence of central bulges at $z \sim 0.8$ -2.6 does not mainly depend on the main galaxy disk having a low angular momentum.

The black filled circles in the right panel of Figure 2 denote all SFGs in our initial disk sample but with $v_{rot}/\sigma_0 < 2$ (see Section 2.1), that is, SFGs with relatively low rotational support and including dispersion-dominated SFGs. These SFGs appear to form the low tail of the specific angular momentum distribution, and have predominantly low stellar masses and small effective radii (< 3 kpc), as pointed out earlier by Newman et al. (2013). These low observed specific angular momenta are not the result of systematically lower inclinations (face-on disks) of the dispersion-dominated SFGs as compared to the entire 'full' sample, nor are they correlated with larger beam smearing corrections, both of which might suggest a systematic underestimate of the intrinsic rotation velocities. We thus conclude that the low specific angular momentum of dispersion-dominated systems most likely is an intrinsic property.

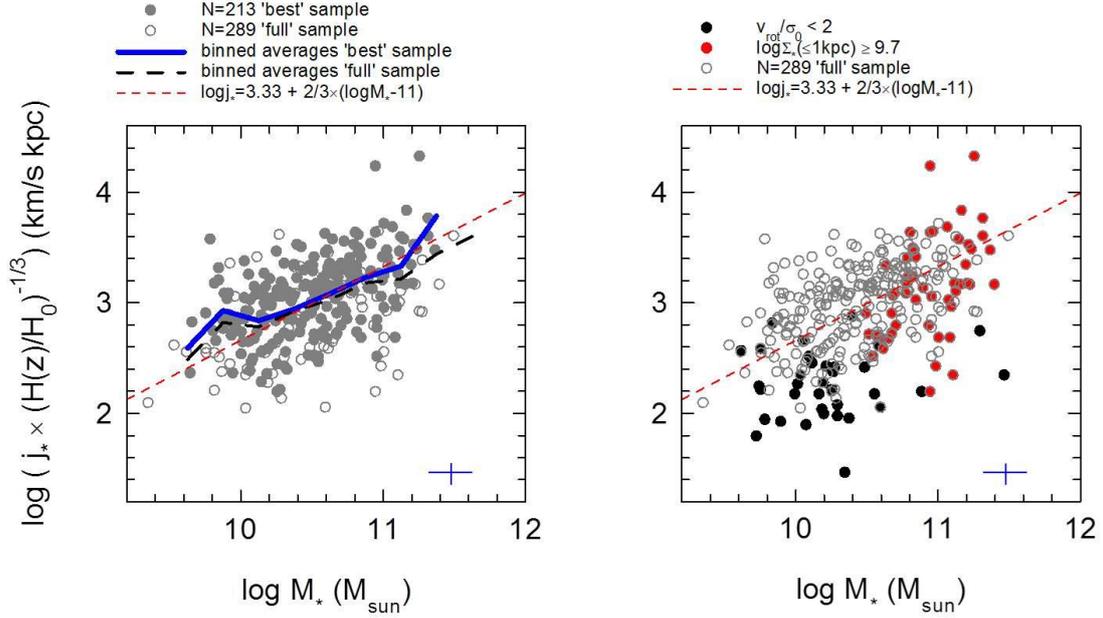

**Figure 2.** Left panel: specific angular momentum in the stellar disk (adopting equation 5) as a function of stellar mass, after removal of the redshift dependence in equation (4) (multiplying the observed $j_*$ with $H(z)^{1/3}$) for the disks of the 'full' (open circles) and the 'best' (filled circles) disk samples. The typical uncertainty is shown as a cross in the lower right. The dashed red line is the best linear fit of slope 2/3 (equation 4) to the 'full' sample (zero point -3.33). The thick dashed black and solid blue lines mark the average trend-line of the 'full' and 'best' samples in bins of 0.25 dex in $\log(M_*)$. Right panel: Same symbols as the left panel, but now emphasizing as filled red circles SFGs in the upper 25% percentile of central stellar surface density $\Sigma_*(R \leq 1 \text{ kpc}) > 10^{9.7} M_\odot \text{kpc}^{-2}$ (equivalent to $n_S > 3$), and as black filled circles all galaxies from our disk sample with $v_{rot}/\sigma_0 < 2$, that is, SFGs with low rotational support and dispersion-dominated SFGs.

### *3.1.2. Comparison to Observations at z~0 and to Recent Simulations*

After elimination of the redshift and stellar mass dependencies of the specific angular momentum, by multiplying $j_d$ with $H(z)^{1/3} \times M_*^{-2/3}$ (see equation 4), we compare in the left panel of Figure 3 the high-$z$ SFGs of the 'full' sample to the late-type disks (filled red squares), early-type disks (brown triangles), and E/S0 galaxies (black crossed squares) in the local Universe from the compilation by Fall & Romanowsky (2013). The observed angular momentum distributions of high-$z$ and $z \sim 0$ disk galaxies (which include Sa galaxies) ***are in excellent agreement***. The tail of low angular momentum SFGs at high-$z$, including the dispersion-dominated SFGs, stretches down to 0.8 dex below the trend-line, in the same region occupied by the $z \sim 0$ spheroidal galaxies in the Fall & Romanowsky (2013) compilation. This agreement of low- and high-$z$ galaxies in terms of their angular momentum distributions is by no means trivial, given that high-$z$ SFGs are 4-7 times more gas-rich in terms of their central (molecular) gas reservoirs (Tacconi et al. 2010, 2013; Daddi et al. 2010), experience more frequent perturbations from (dissipative) minor and major mergers (Fakhouri & Ma 2008; Genel et al. 2009; Rodriguez-Gomez et al. 2015), and exhibit much more commonly powerful galactic outflows that might alter the angular momentum distribution of the disk (Übler et al. 2014).

The observed average specific angular momenta as a function of stellar mass in the left panel of Figure 3 are also ***in impressive agreement with the recent generation of hydrodynamical simulations*** (Illustris, Genel et al. 2015; EAGLE, Zavala et al. 2015). We show in the left panel of Figure 3 the predicted specific angular momenta in the Illustris simulation as green ($z = 2$), blue ($z = 1$) and magenta ($z = 0$) lines (Genel et al. 2015; S. Genel, private communication), which agree quite well with both low- and high-$z$ data (but fall on average ~ 0.1-0.2 dex below the average trend-lines of the data).

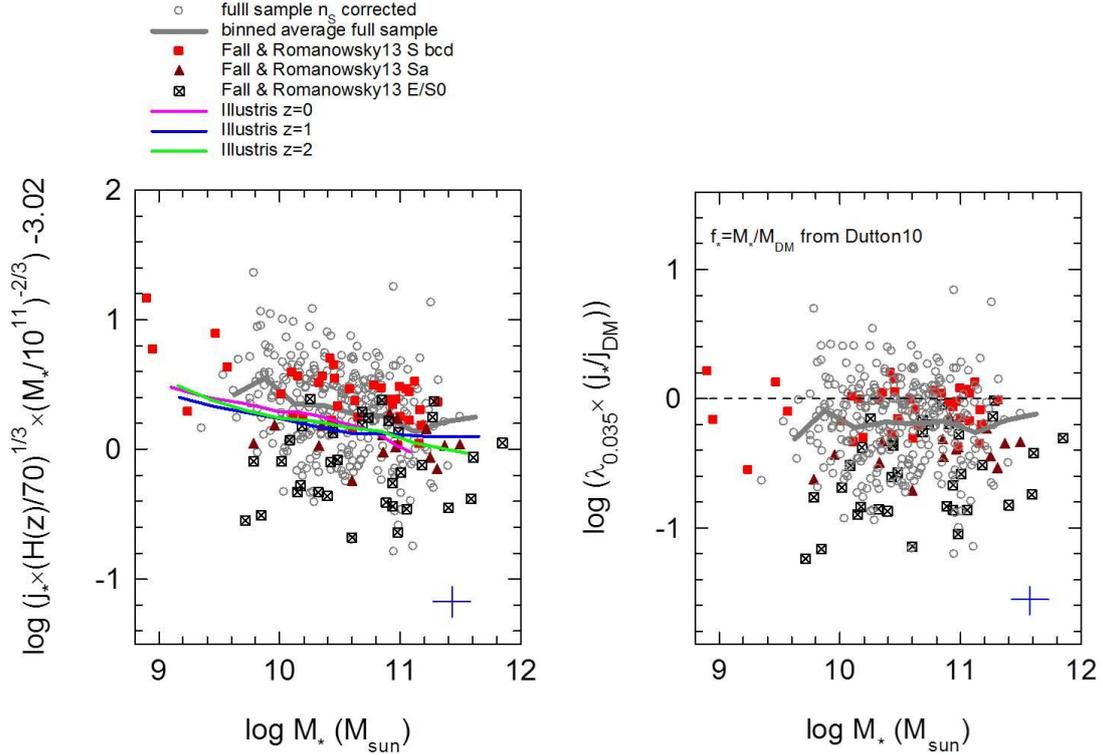

**Figure 3.** Left panel: distribution of the observed specific angular momenta of the $z =$ 0.8 - 2.6 SFGs in the 'full' sample (grey open circles, and thick grey solid trend-line of the binned averages, both as in Figure 2), after multiplying the data in Figure 2 with $M_*^{-2/3}$ in order to remove the stellar mass dependence, as well as the redshift dependence (see equation 4). Filled red squares denote the $z = 0$ late-type disks (Sbcd), brown filled triangles the early-type disks (Sa), and black crossed squares the spheroidal galaxies (S0,E) in the compilation of Fall & Romanowsky (2013). Green, blue, and magenta lines are the predictions of angular momenta of star forming galaxies from the Illustris hydrodynamical simulation (Genel et al. 2015; S. Genel, priv.comm.). Right panel: Stellar mass dependence of $\log(j_d/j_{DM})$ in the high-$z$ and low-$z$ data of Figure 2 (same symbols as in the left panel), with the assumption of $\lambda =$ 0.035, after removing the $f_*(M_*)$ dependence with the fitting function of Dutton et al. (2010; similar to Moster et al. 2013).

### 3.1.3 Eliminating $f_*$

Following the motivation of Section 1, our next goal is now to use the data and equation (4) to gain insights on the ratio of baryon to DM specific angular momenta in our galaxies. For this purpose, we need to eliminate the dependency on the function $f_*(M_*)$. To do so, and again following Romanowsky & Fall (2012), we use the fitting function of Dutton et al. (2010), who empirically derived the stellar baryon fractions of local Universe galaxies from a combined analysis of stellar kinematics, weak lensing, and abundance matching. The Dutton et al. (2010) fitting function (for late-type star forming galaxies at $z = 0$) is given by

$$f_*(M_*)_{D10} = 0.29 \times \left(\frac{M_*}{5 \times 10^{10} M_\odot}\right)^{0.5} \times \left[1 + \left(\frac{M_*}{5 \times 10^{10} M_\odot}\right)\right]^{-0.5} \quad (6).$$

Dutton et al. (2010) also gave a similar fitting function for early-type galaxies. A similar correction function (in terms of mass dependence and zero point) is obtained from the Moster et al. (2013) or Behroozi et al. (2013a,b) abundance matching. We decided to take the Dutton et al. (2010) fitting function in equation (6), since it extends to $\log(M_*/M_\odot) > 11$ where we have many SFGs in our sample.

The right panel of Figure 3 shows the results of applying this correction, which should now give a quantitative estimate of the ratio of disk to dark matter specific angular momentum as a function of stellar mass, all again under the simplifying assumption of a constant dark matter angular momentum parameter. With these assumptions we find $< \log(j_d/j_{DM}) > -0.2$ dex, ***independent of stellar mass*** between $\log(M_*/M_\odot) = 9.5$ and 11.6, and including the tail of lower angular momentum galaxies discussed before. Leaving out these extreme outliers by selecting the 'best' sample yields an average of -0.09. For comparison, the average of all Fall & Romanowsky (2013) star forming spiral galaxies (including Sa types) is -0.17, while the late-type systems (Sbcd) have an average of -0.03. We conclude that star forming galaxies between $z = 0$ and 2.6 plausibly have on average retained between 60% and 90% of their dark matter specific angular momentum in their main baryonic disk.

## *3.2 Angular Momentum Parameter*

Based on the analysis in Section 2.2 and in the Appendices A and B, we now have several estimates of λ parameters, or more precisely of $\lambda \times (j_d/j_{DM})$, for rotation-dominated SFGs of the 'full' and 'best' samples. Uncertainties of the individual measurements range from ±0.06 to ±0.33 dex in logarithmic units, with a median of ±0.2 dex.

Figure 4 (left panel) shows the distribution function of these $\lambda$ parameters for our 'primary' NFW MC modeling (method 1 in Section 2.2, described in detail in Appendices B4 and B5), and assuming pure (baryonic) exponential disks. This distribution (plotted as histogram) is well fitted by a log-normal function of intrinsic dispersion ~ 0.17 dex in log $\lambda$, after subtracting the measurement uncertainties in quadrature from the measured dispersion of the distribution. Taking the 'full' and 'best' samples with converged NFW models yields error-weighted averages of $\langle\lambda\times(j_d/j_{DM})\rangle$ = 0.039 and 0.041, respectively (first two rows of Table 1). An unweighted Gaussian fit to the modeling results for the 'full' sample in Figure 4 yields $\langle\lambda\times(j_d/j_{DM})\rangle$ = 0.032 (third row of Table 1). The intrinsic dispersion of these distributions varies between $\leq$ 0.16 and 0.19 dex in log $\lambda$.

Red triangles denote the same NFW MC modeling methodology, for pure exponential disks but this time *with* adiabatic contraction, as described by Mo et al. (1998). The resulting distribution again is log-normal with a similar dispersion, but with a greater mean, $\langle\lambda\times(j_d/j_{DM})\rangle$ = 0.044 to 0.071, depending on whether we use an unweighted or weighted estimator (fourth row of Table 1). Green circles again denote the same NFW MC fitting, without adiabatic contraction (as for the black shaded diagram), but now implementing individual corrections to $\lambda$ values for surface density distributions deviating from the pure exponential distributions assumed so far, and obtained from free $n_S$ fits to the rest-frame optical continuum distributions for each SFG (see Appendix B.7). The resulting mean, given in the fifth row of Table 1, is $\langle\lambda\times(j_d/j_{DM})\rangle$ = 0.051. The intrinsic dispersions of these distributions range between $\leq$ 0.16 and 0.22 dex in log $\lambda$.

In the right panel of Figure 4 we compare our 'primary' NFW MC modeling (again depicted as histogram) with the other, simpler methods discussed in Section 2.2 and Appendix B. All result in smaller mean values of $\langle\lambda\times(j_d/j_{DM})\rangle$. Red triangles denote the results for the 'best' disk sample if instead of the NFW MC modeling the simplest, isothermal model (equation (B1)) is adopted. This yields $\langle\lambda\times(j_d/j_{DM})\rangle$ ~ 0.02 with intrinsic dispersion $\sigma(\log \lambda)$ ~ 0.22 (sixth row of Table 1). Green circles denote the distribution of $\lambda$ values that is obtained from equation (B9), and when $M_{DM}$ = $(M_*+M_{molgas})_d/m_d$ values are estimated by inverting the Moster et al. (2013) fitting functions to infer the halo mass from the stellar mass. The resulting $\lambda$ distribution has $\langle\lambda\times(j_d/j_{DM})\rangle$ = 0.016 and $\sigma(\log \lambda)$ = 0.3 (row 7 in Table 1). Finally, if in equation (B9) a constant value of $m_d = 0.05$ is adopted, motivated by the average $\langle m_d \rangle$ value obtained from our primary NFW MC method, the resulting distribution is shown by the cyan squares and has $\langle \lambda \times (j_d/j_{DM}) \rangle = 0.023$ and $\sigma(\log \lambda) \sim 0.24$. If in all these three cases the 'full' instead of the 'best' sample is used, the centroid is similar, but the scatter is larger.

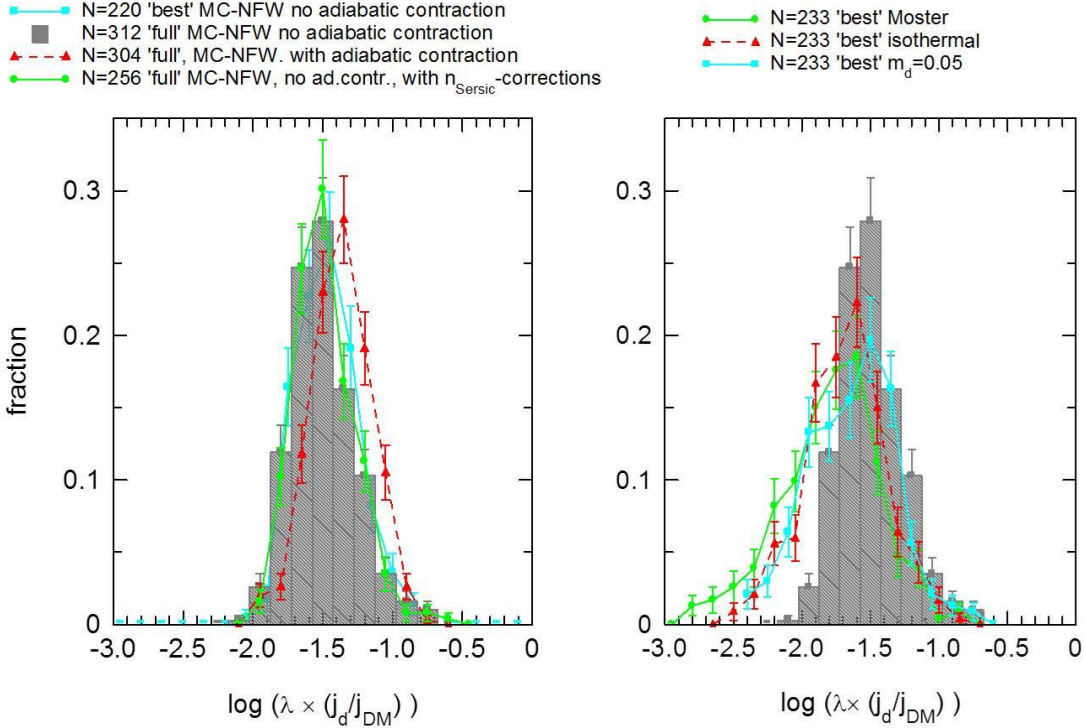

**Figure 4.** Distribution of inferred $\lambda \times (j_d/j_{DM})$ parameters. Left: Distribution for all 312 $z = 0.8 - 2.6$ rotation-dominated SFGs in the 'full' sample with converged NFW MC modeling, under the assumption of no adiabatic contraction of the dark matter halo (histogram, with $1\sigma$ Poissonian uncertainties, section B.5). The cyan squares, the red triangles and the green circles denote the same NFW MC modeling, but this time with the 'best' sample, with the 'full' sample but including adiabatic halo contraction, as well as without adiabatic contraction but including the corrections for varying Sérsic indices, respectively (Appendix B.7). Right: For comparison with the histogram in the left panel we show three other, simpler estimates of the $\lambda$ distribution. Cyan squares denote the distribution of $\lambda$ values that is obtained for the 'best' disk sample if instead of the NFW MC modeling the simple assumption $M_{DM} = (M_* + M_{molgas})_d/m_d$ is made and $\lambda$ is estimated from equation (B9) with $m_d = 0.05$. If instead the full sample is used, the centroid is similar, but the scatter is larger. When using the Moster et al. (2013) fitting functions to infer the halo mass from the stellar mass (and then obtaining $\lambda$ from equation B9 again), the resulting $\lambda$ distribution is given by green circles. Finally, the results of the simplest, isothermal model are marked by red triangles (equation (B1)).

**Table 1. Summary of Determinations of $\langle\lambda\times(j_d/j_{DM})\rangle$**

| Sub-sample | N | $\langle\lambda\times(j_d/j_{DM})\rangle$[1] | $\delta_2 (\langle\lambda\times(j_d/j_{DM})\rangle)$[2] | $\sigma(\log\lambda)$[3] |
|---|---|---|---|---|
| NFW MC, 'full', error weighted no adiab.contraction | 312 | 0.039 | 0.003 | 0.17 |
| NFW MC, 'best', error weighted no adiab.contraction | 220 | 0.041 | 0.004 | 0.19 |
| NFW MC, 'full', equal weight no adiab.contraction | 312 | 0.032 | 0.003 | ≤0.16 |
| NFW MC, 'full', error weighted with adiab.contraction | 304 | 0.071 | 0.004 | ≤0.16 |
| NFW MC, 'full', error weighted no adiab.contraction with Sersic corrections | 256 | 0.051 | 0.0044 | 0.22 |
| Isothermal, 'best', equal weight | 233 | 0.020 | 0.002 | 0.22 |
| 'Moster', 'best', equal weight | 233 | 0.016 | 0.002 | 0.3 |
| $m_d$=0.05, 'best', equal weight | 233 | 0.023 | 0.002 | 0.24 |
| **final overall average** | | **0.037** | **0.004 (stat)** **0.018 (syst)** | **0.2** |

1) weighted mean of log λ distribution
2) twice the uncertainty of the mean of weighted log λ distribution
3) dispersion of weighted log λ distribution, after subtraction in quadrature of the median measurement error

In our view the 'primary' NFW MC fitting constitutes the best technique for estimating halo masses and the corresponding angular momentum parameters because it makes the fewest assumptions and takes into account all the relevant measured properties and their observational uncertainties. However, there is a significant degeneracy between λ and $M_{baryon}/M_{DM}$ in this technique (see Figure 11 in Appendix B: log λ ~ –0.8 + 0.58×log($M_{baryon}/M_{DM}$) ). Looking at Table 1 the agreement of the results of the different methods is quite encouraging, keeping in mind the substantially different assumptions involved in each of the entries. In terms of the mean, the results of the different methods scatter both to larger, as well as to smaller, values than our 'primary' NFW MC modeling. In terms of scatter, the 'simpler' methods yield somewhat larger scatter, as might have been expected. The comparison in Table 1 gives a good indication of the systematic uncertainties, which are significantly larger than the formal fit uncertainty in each of the entries in Table 1. Dispersion-dominated objects, with low inferred specific angular momenta (see

Section 3.1 and Figure 2), were excluded in the $\langle\lambda\times(j_d/j_{DM})\rangle$ distributions discussed here, which may therefore be biased in their mean and scatter. However, in our data sets these objects represent only about 10% of the size of our 'full' sample such that the bias is small compared to other uncertainties from the assumptions or methodology described above.

We adopt as the final mean inferred parameter $\langle\lambda\times(j_d/j_{DM})\rangle = 0.037$, with a small statistical uncertainty of ±0.004 and a dominant systematic uncertainty of ±0.018.

We now explore the correlation of the inferred angular momenta with redshift, and with various disk and halo properties. The most important parameter correlations are shown in Figures 5 and 6. In each panel we show the 'best' (filled blue circles) and 'full' (filled and open blue circles) samples, and give the equal weight, trend-line of binned averages (thick continuous grey curve) as well as the best, error weighted linear fit in log-log space (i.e. a power law) to the 'full' sample (dashed red line). The fit parameters (zero point and slope) for these weighted fits are listed, as are the correlation coefficients (R). In Figure 5, the panels at the top and bottom show the strongest and weakest correlations, respectively. In Figure 6, the plots are sorted by increasing correlation strength from left to right.

Figure 5 shows that the inferred $\lambda\times(j_d/j_{DM})$ distribution, its centroid and dispersion do not depend much on redshift in the interval covered by our data, nor on the stellar or halo masses, nor on the central concentrations of the galaxies, as traced by the Sérsic index of the rest-frame light distribution, nor on the stellar surface density in the central 1kpc.

Figure 6 shows the three strongest correlations (with slopes differing from zero at the 10-15$\sigma$ level) between $\lambda\times(j_d/j_{DM})$ and the disk scale length, the stellar surface density within the effective radius, and the rotation velocity at the effective radius. The strongest correlation is between $\lambda\times(j_d/j_{DM})$ and disk radius. This is interesting as the specific angular momentum is essentially a product of rotational velocity and scale radius. Equation (3), however, shows that for galaxies with a given virial radius or virial mass, one would expect a linear correlation of $\lambda\times(j_d/j_{DM})$ with disk radius. We also find significant correlations with the stellar surface density within the half-light radius and with the rotation velocity at $R_{1/2}$ (middle and left panels of Figure 6). Since stellar mass and spin parameter are not significantly correlated, the correlation between stellar surface density and $\lambda$ (dlog$\Sigma_*$/d$\lambda \sim -3.4$) is probably largely induced by the strong correlation between disk radius and $\lambda$ ($dR_d/d\lambda \sim 1.33$) since $\Sigma_* \sim M_*/R_d^2$. The anticorrelation between $\lambda \times (j_d/j_{DM})$ and $v_{rot}$, though it seems counterintuitive, follows from Equations (1) and (3) as described in Appendix B.1.

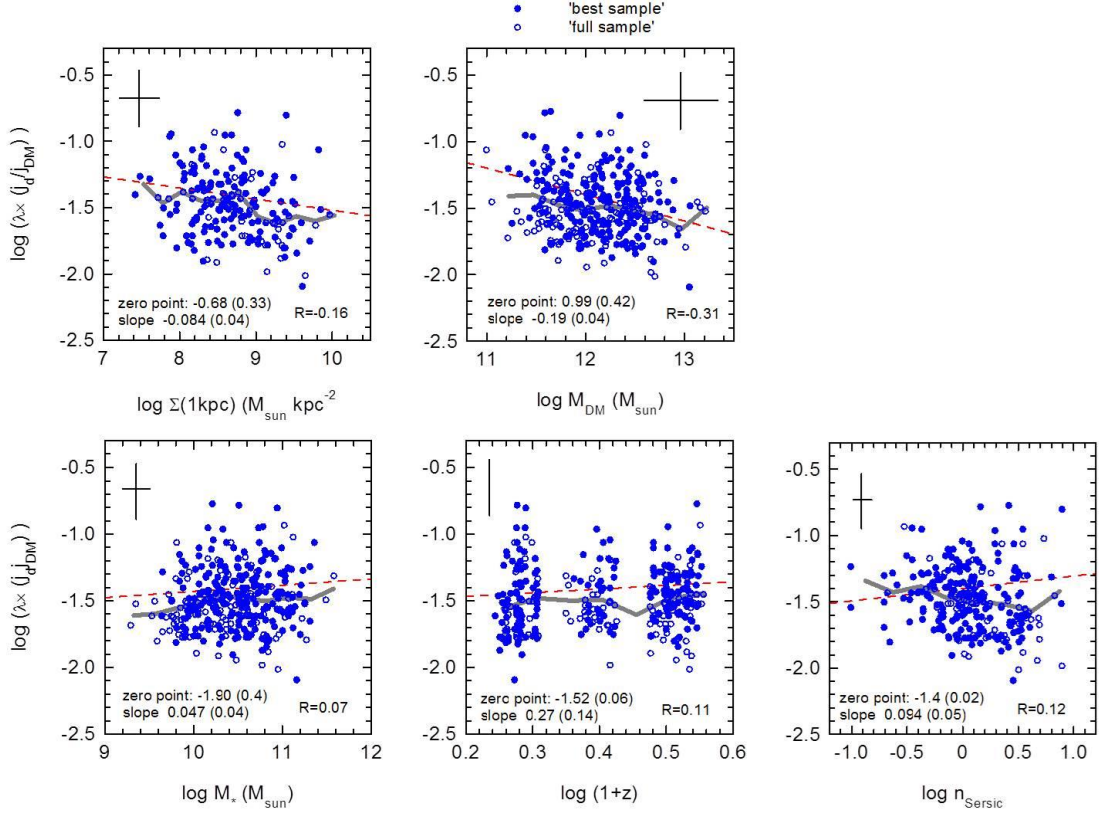

**Figure 5.** Dependence of the inferred $\lambda \times (j_d/j_{DM})$ parameters from the NFW MC modeling for the 'best' (filled blue circles) and 'full' (filled plus open blue circles) samples as a function of (from top left to bottom right) stellar surface density in the central 1 kpc, halo mass, stellar mass, redshift, and Sérsic index of the rest-frame $R$-band continuum light. In each panel large crosses denote the typical uncertainties, thick grey curves denote the trend line of binned, equal weight averages of the 'full' sample data, and the red dotted line is the best linear, error-weighted fit in log-log space (i.e., a power law) to the 'full' sample data. Zero points and slopes, and their 1$\sigma$ uncertainties, along with the correlation coefficients (R), are listed as well. The panels in each row are ordered in ascending correlation strength from left to right (with the two stronger correlations in the top row). All trends explored in this Figure show little, if any, correlation, and have a slope consistent with zero at the 2$\sigma$ level or less except for the trend with $M_{DM}$, for which the slope differs from zero at the 4.8$\sigma$ level).

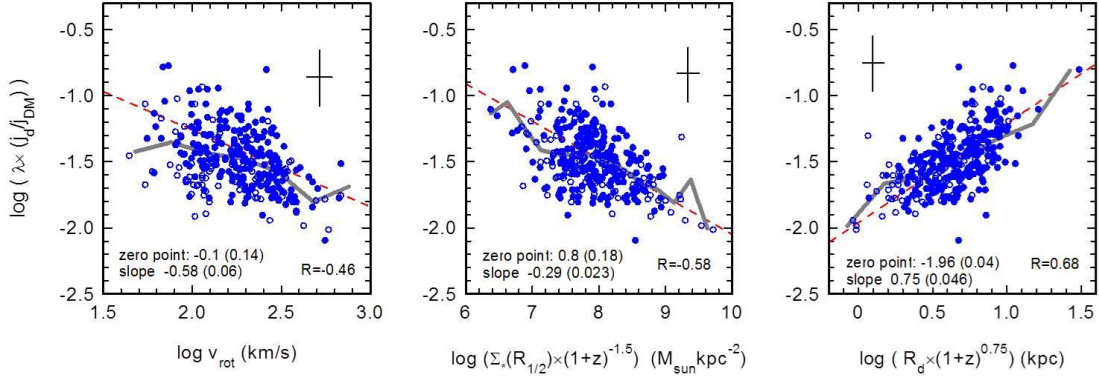

**Figure 6.** Dependence of the inferred $\lambda \times (j_d/j_{DM})$ parameters from the NFW MC modeling for the 'best' (filled blue circles) and 'full' (filled plus open blue circles) samples as a function of (from left to right) rotation velocity at $R \sim R_{1/2}$, stellar surface density within $R_{1/2}$ and $R_d/(1+z)^{-0.75}$. The nomenclature is the same as in Figure 5. The three plots in this Figure show the strongest parameter correlations among the trends explored in Section 3.2 (with slopes differing from zero at the 10–15σ level).

In the $\lambda$-$R_d$ and $\lambda$-$\Sigma$ correlations in Figure 6 we have removed the mean redshift dependence of the disk sizes. The fitting function for the dependence of *population averaged* disk scale length on redshift for star forming galaxies obtained by van der Wel et al. (2014a) from CANDELS near-IR HST imagery[3] gives

$$A(z) = (1+z)^{-0.75} \qquad (7).$$

Dividing the observed disk half-light radii by equation (7) then yields the correlation in Figure 6, which has a scatter of ±0.17 dex. One can turn this finding around and use this correlation as a one parameter estimator of angular momentum parameters of SFGs without kinematic data,

$$\log\left[\lambda \times \left(\frac{j_d}{j_{DM}}\right)\right] = -1.96 \, (\pm 0.04) + 0.75 \, (\pm 0.046) \times \log\left(\frac{R_d}{A(z)}\right) \qquad (8).$$

A similar relation of slightly larger scatter and somewhat poorer correlation is obtained without the redshift correction $A(z)$. In that case the zero point and slope are -1.71(±0.03) and +0.68 (±0.05).

---

[3] Equation (7) is based on the redshift evolution of the linear $\log(R_{1/2})$ - $\log(M_*)$ relationship of SFGs from the same 3D-HST/CANDELS reference population discussed in Section 2.1; Figure 1 shows that the size distribution of our IFU sample and this reference SFG population is essentially identical over the same redshift and mass range, when excluding galaxies well below the main sequence, such that the same redshift evolution should apply for our galaxies.

For the SFGs with highest stellar surface densities, $\log(\Sigma_*(R \leq R_{1/2})\ [M_\odot\ \mathrm{kpc}^{-2}]) \geq 9$, the spin parameters are about half ($<\lambda \times (j_d/j_{DM})> \sim 0.018$) that of the median value of all SFGs. Including the Sérsic index corrections discussed in Appendix B.7 makes no significant difference, neither for the average $\lambda \times (j_d/j_{DM})$ value, nor for the (lack of) mass dependence. The correlations of $\lambda \times (j_d/j_{DM})$ with $R_d$ and $\Sigma_*$ become slightly flatter, but within the uncertainties of the fits shown in Figure 6.

In summary of this section there are three main conclusions. First, we *find a near-universal, log-normal distribution of $\lambda \times (j_d/j_{DM})$, whose centroid (0.037) and dispersion (0.2 in $\log \lambda$) are very similar to that inferred for the cold dark matter component* as determined from CDM simulations. The *stellar surface density within the half-light radius and rotation velocity exhibit a significant negative correlation with $\lambda \times (j_d/j_{DM})$. More compact and denser SFGs have lower values of $\lambda \times (j_d/j_{DM})$*, either because they had initially smaller dark matter $\lambda$ values, or because a fraction of the baryons suffered significant angular momentum loss between the halo and circum-nuclear scale. This result is in very good agreement with previous work in disk-dominated galaxies in the local Universe (Fall 1983; Dutton & van den Bosch 2012; Romanowsky & Fall 2012; Fall & Romanowsky 2013; Courteau & Dutton 2015). Third, *the lack of correlation between $\lambda \times (j_d/j_{DM})$ and $\Sigma_*(1\mathrm{kpc})$ (or $n_S$) suggests that the central bulges are decoupled from the kinematic properties of the outer disk to which our data are sensitive*.

### 3.3. Baryon to Dark Matter Mass Ratios

Our NFW MC modeling gives the ratio of the baryonic and stellar masses in the disk to the mass of the dark matter halo. We thus can compare our results of the ratio of $M_*/M_{DM}$ as a function of $M_{DM}$, obtained with kinematic data, with the totally independent results from abundance matching (Conroy & Wechsler 2009; Moster et al. 2010, 2013; Behroozi et al. 2010, 2013a). The resulting dependence of the ratio $M_*/M_{DM}$ as a function of halo mass is shown in Figure 7, again for our 'full' and 'best' samples with converged NFW MC models. For comparison we show the abundance matching results at $z \sim 1$ and 2 (Moster et al. 2013; Behroozi et al. 2013a).

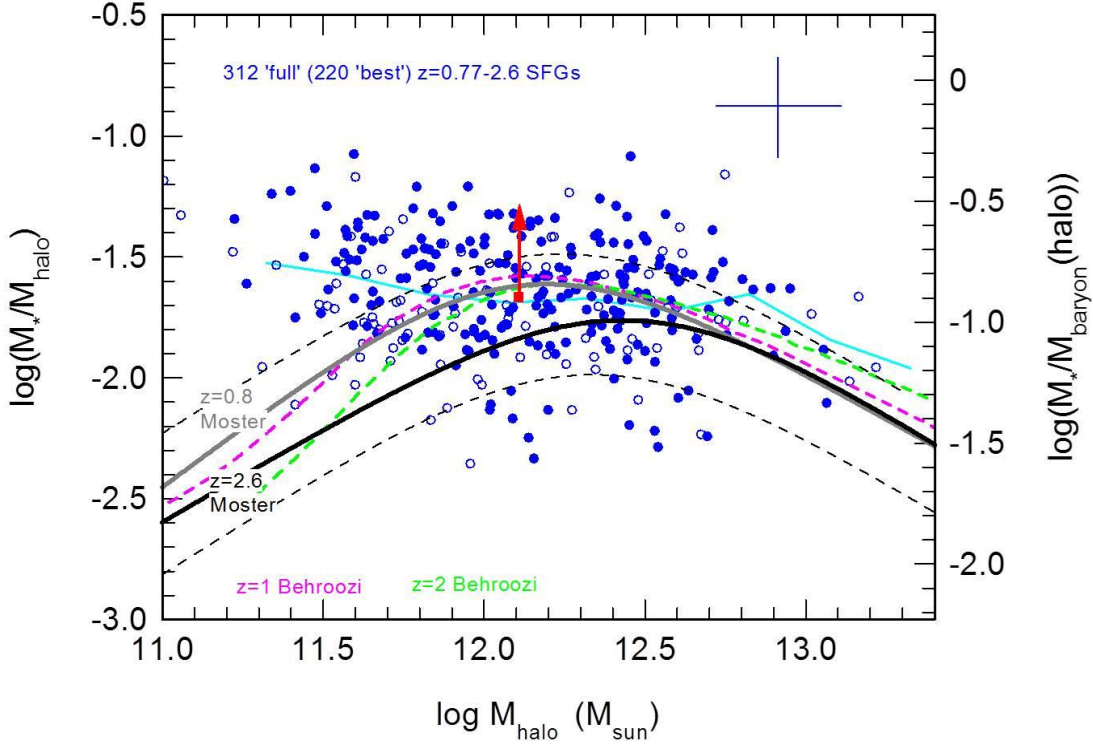

**Figure 7.** Stellar to total halo mass ratios as a function of $M_{DM}$ for our disks of the 'full' (open and filled blue circles) and 'best' (filled blue circles) samples from our NFW MC modeling. The blue cross in the upper right denotes the typical 1σ uncertainties. The continuous cyan line denotes the average trend of the 'full' sample in bins of 0.25 dex in $\log(M_{DM})$. The red filled square marks the median value of our measurements. The red arrow denotes the average correction from stellar to baryonic disk masses, including the molecular gas contribution ($<M_{gas}/(M_{gas}+M_*)> \sim 0.58$). The right vertical axis denotes the ratio of the stellar disk mass to the total baryon mass, adopting a cosmic baryon to dark halo fraction of 0.17. The thick grey and black curves give the fitting function obtained by Moster et al. (2013) from rank ordered abundance matching of stellar mass functions and dark matter halo simulations for $z = 0.8$ and $z = 2.6$, respectively. The black dashed curves denote the ±1σ uncertainty of the combined $z = 0.8$ and $z = 2.6$ Moster et al. (2013) models. Magenta and green thick dashed lines mark the $z = 1$ and $z = 2$ abundance matching results of Behroozi et al. (2010, 2013a).

The red up-arrow in Figure 7 denotes the average correction from stellar to total baryonic mass in the disk. We find that $<M_*/M_{DM}> \sim 0.022$ and $<m_d> = <(M_* + M_{molgas})/M_{DM}> \sim 0.052$ such that for halos near $10^{12}\ M_\odot$, 13% and 31% of the cosmologically available baryons are in the stellar and baryonic disk, respectively. For NFW MC modeling including adiabatic contraction the value for $<m_d>$ would increase to about 0.1, with scatter twice as large. We find little dependence of $M_*/M_{DM}$ on halo mass in the range $\log(M_{DM}/M_\odot) \sim 11.5$-13.2 sampled by our measurements.

These results are in reasonable agreement with those obtained from the abundance matching technique. Our stellar to dark matter mass ratios are on average 35% larger than predicted by Moster et al. (2013), and comparable to Behroozi et al. (2013a). These differences are well within the uncertainties, as both the abundance matching and our kinematic methods have substantial systematic uncertainties (as shown in Figure 7)[4].

In the abundance matching results, the maximum stellar to halo mass ratio $(M_*/M_{DM})_{peak}$ at $M_{DM} \sim 10^{12} M_\odot$ and at $z \sim 2$ is about the same (Behroozi et al. 2013a,b), or 0.6 times (Moster et al. 2013) that at $z \sim 0$. Because the gas fractions are much higher at $z \sim 2$ (Tacconi et al. 2010, 2013; Sargent et al. 2014; Genzel et al. 2015; Béthermin et al. 2015), the **baryon** to DM mass ratio, $M_{baryon}/M_{DM}$ for a $M_* = 5 \times 10^{10} M_\odot$ SFG located on the main-sequence is then about 1.4 times (Behroozi) and 2.3 times (Moster) larger at $z = 2$ than at $z = 0$. Together with our somewhat larger $M_*/M_{DM}$ ratios as compared to the abundance matching method, our data suggest that the peak baryon to dark matter mass ratio at $z \sim 2$ is about 2-3 times larger than at $z=0$.

The main issue in the case of our NFW MC modeling is the fact that on the 2-7 kpc scale of the disk sampled by our Hα kinematics, most of the mass is due to the baryons, with an average dark matter fraction in the disk of 25±15%, such that the extrapolation to the halo scale is naturally uncertain by $\geq 0.2$ dex. We refer to Förster Schreiber et al. (2009) and S. Wuyts et al. (2016) for a more in depth discussion of the evidence that the disks at $z \sim 1$-2.6 are strongly baryon dominated.

---

[4] Our data do not exhibit a decrease in $M_*/M_{DM}$ towards lower halo masses, in contrast to the expectations from abundance matching. However, this regime is dominated by $\log(M_*/M_\odot) \lesssim 10.5$ galaxies, where our current kinematics sample starts to become significantly incomplete with respect to the underlying galaxy population (see Figure 1). This potentially interesting trend will be pursued in the future, when the sample from our on-going KMOS[3D] survey becomes larger and more complete at lower stellar masses.

## *3.4. Comparison to Theoretical Models of Galactic Disk Formation by Gas Accretion from the Cosmic Web*

We have shown in section 3.2 that the spin parameter distribution of the near-main sequence star forming galaxies has a mean of $\langle\lambda\times(j_d/j_{DM})\rangle \sim 0.037$ and a dispersion of $\sigma_{\log \lambda} \sim 0.2$, which is in agreement with the distribution of dark halo spins of virialized dark matter halos as inferred from cosmological simulations (Bullock et al. 2001a; Maccio et al. 2008). This finding is in agreement with previous simple analytical models (e.g., Fall & Efstathiou 1980; Mo, Mao & White 1998) that assumed $j_d = j_{DM}$ and with recent numerical simulations (e.g. Fiacconi et al. 2015; Genel et al. 2015; Teklu et al. 2015; Pedrosa & Tissera 2015; Danovich et al. 2015). Although expected in the light of these studies, we still consider this finding by no means trivial and quite surprising. We note also that previous studies were devoted to present-day galaxies whereas we focus here on the high-redshift universe where filamentary gas accretion from the cosmic web is likely to dominate the growth of galaxies. High-resolution numerical simulations of galaxy formation by filamentary accretion indeed find that gas entering the virial radius at a given time has 2-3 times more angular momentum than the corresponding dark matter that is being accreted at that time (Kimm et al. 2011; Pichon et al. 2011; Stewart et al. 2011, 2013; Danovich et al. 2012, 2015; Teklu et al. 2015). Danovich et al. (2015) recently investigated in detail the angular momentum evolution of gas while it settles into the galactic disk of high-$z$ galaxies. They identified four characteristic phases of angular momentum exchange that in the end 'conspire' such that the gas has the same net specific angular momentum as its dark halo when it enters the disk region and settles into centrifugal equilibrium. In this case, galactic disks should indeed reflect the spin parameter distribution of dark halos, in agreement with our empirical results. The origin of this remarkable 'conspiracy' is however not clear.

Additional processes within the star forming disks could in principle change $j_d$ substantially, destroying a correlation with $j_{DM}$. Galactic disks are strongly evolving internally due to viscous accretion, leading to angular momentum redistribution and gas inflow within the disk plane towards the galactic center (Noguchi 1999; Immeli et al. 2004a,b; Krumholz & Burkert 2010; Forbes et al. 2014a,b; Bournaud et al. 2014). This process would not change $j_d$. However, large amounts of angular momentum could then be stored in extended, non-star forming HI envelopes that are not easily detectable. Gas is also ejected by supernova and/or active galactic nuclei (AGN) driven outflows, and it is unlikely that disks at all radii eject the same fraction of gas (see Übler et al. 2014). Still, our results indicate that the ejected gas must have the same specific angular momentum as the dark halo and disk, in order for the spin distribution of the disks not to change relative to the dark halo spin.

## 3.5 Dependence of $\lambda \times (j_d/j_{DM})$ on Surface Density: Nature of Nurture?

From observations, analytic work, and simulations, a number of authors in the last decade have proposed that a fraction of the $z > 1$ SFGs must undergo an internal 'fast compaction' leading to the formation of massive, gas-rich bulges, prior to quenching and transitioning to the passive galaxy population (Tacconi et al. 2008; Cimatti et al. 2008; Hopkins et al. 2009; Engel et al. 2010; Barro et al. 2013, 2014a, 2015, 2016; Lang et al. 2014; Dekel & Burkert 2014; Zolotov et al. 2015; Tacchella et al. 2015a,b, 2016; Wellons et al. 2015; but see van Dokkum et al. 2015 for a contrasting view). Several of the former authors argued that 'wet compaction' (with gas and stars being transported radially inwards) may be triggered by a combination of mergers (major and/or minor; e.g., Tacconi et al. 2008; Cimatti et al. 2008; Hopkins et al. 2009), as well as the 'violent disk instability' acting efficiently in gas-rich galaxies (Noguchi 1999; Immeli et al. 2004a,b; Bournaud et al. 2007, 2014; Genzel et al. 2008; Dekel et al. 2009; Cacciato et al. 2012; Dekel & Burkert 2014; Zolotov et al. 2015). Franx et al. (2008), Bell et al. (2012) and Lang et al. (2014) have shown that the quenched galaxy fraction is a strong function of central velocity dispersion, $\Sigma_*(1kpc)$, $n_S$ and $M_{bulge}$. Is this 'wet' compaction model consistent with our kinematic data?

We think the answer is *yes*, but in an indirect way. Surface density correlates with galaxy baryonic mass. In the left panel of Figure 8 we show the dependence of stellar surface density within the half light radius $\Sigma_*(R_{1/2})$, as well as of the molecular gas surface density within $R_{1/2}$, $\Sigma_{gas}(R_{1/2})$, and of the stellar surface density within the central 1 kpc, $\Sigma_*(1kpc)$, as a function of stellar mass, after removal of the average redshift dependence (as in Figure 6).

All surface densities increase with stellar mass (e.g., $\langle\Sigma_*(R_{1/2})\rangle \sim M_*^{0.54}$, van der Wel et al. 2014a) but the slopes are significantly different. The inferred molecular surface densities increase more slowly than the average stellar surface density, implying lower gas fractions in the higher mass SFGs (see Tacconi et al. 2013; Saintonge et al. 2013; Sargent et al. 2014; Genzel et al. 2015)[5]. In turn the stellar surface densities in the central 1 kpc increase still faster with mass than the galaxy averaged stellar surface densities. In agreement with Barro et al. (2015), we consider this finding a strong argument in favor of the ***internal growth of central mass concentrations (bulges)*** during the evolution of the SFGs along the main-sequence. For the 3D-HST reference sample, $\Sigma_*(1\text{kpc})$ becomes comparable to the surface density of massive quenched galaxies for $\log(M_*/M_\odot) \geq \log(M_S/M_\odot) \sim 10.9$ at $z \sim 2\text{-}3$, and $\log(M_*/M_\odot) \geq 10.6$ at $z \sim 1$ (e.g., Lang et al. 2014; Barro et al. 2015, 2016; van Dokkum et al. 2015). The compact massive SFGs at the dense tip of the trend in Figure 8 were called 'blue nuggets' by Barro et al. (2013, 2014a,b), which tend to have cuspy stellar distributions ($<n_S> \sim 2\text{-}4$). Their bulge to total mass ratios can reach $<M_{bulge}/M_*> \sim 0.5$.

The right panel of Figure 8 shows the inferred angular momentum parameter as a function of the same three measures of surface density. The value of $\lambda \times (j_d/j_{DM})$ decreases with the galaxy wide surface densities, gas and stars, in similar measure, but $\lambda \times (j_d/j_{DM})$ does not, or only weakly, depend on $\Sigma_*(1\text{kpc})$, as we had pointed out in section 3.2 (Figures 5 and 6).

We interpret this finding in the following way. If the formation of the central mass concentrations was mainly due to 'nature' (small $\lambda$ or small $(j_d/j_{DM})$ of the entire disk), then we should see a strong correlation between $\lambda \times (j_d/j_{DM})$ and all tracers of surface density/size. Such a strong correlation of all three tracers would, for instance, be expected if the dominant channel for compaction is major mergers, as they tend to re-distribute angular momentum within the entire galaxy merger remnant (Mihos & Hernquist 1996). The fact that we are observing a strong correlation between the angular momentum parameter (sensitive mainly to the kinematic properties of the outer disk) and the galaxy-wide surface densities, but not with the surface density of the 'compacted' bulge/nucleus suggests to us that ***the main channel of compaction is a galaxy-internal process (or processes), such as radial transport in the disk***

---

[5] A caveat is that our gas masses are based on applying in reverse scaling relations between molecular gas masses and rest-frame optical/UV data. The latter are very sensitive to extinction. If there were highly extincted nuclear starbursts triggered by compaction, they probably would not easily be detected by these data, and instead high resolution sub-mm/mm observations are required.

*instability, or less perturbative minor mergers*. Van Dokkum et al. (2015) have brought forward a slightly different view that the formation of central bulges is a result of the inside-out growth of galaxies as a function of cosmic time ( $d\log R_{1/2} \sim -0.75 \times d\log(1+z) + 0.23 \times d\log M_*$, van der Wel et al. 2014a). In this scenario quenching occurs once a star forming galaxy crosses a threshold in central density or velocity dispersion. Galaxies that reached that threshold earlier in time formed more compact quenched descendents. While the final verdict is still unclear, we favor at present the internal compaction model during mass growth along the main sequence, as brought forward by Dekel & Burkert (2014), Barro et al. (2015), and others.

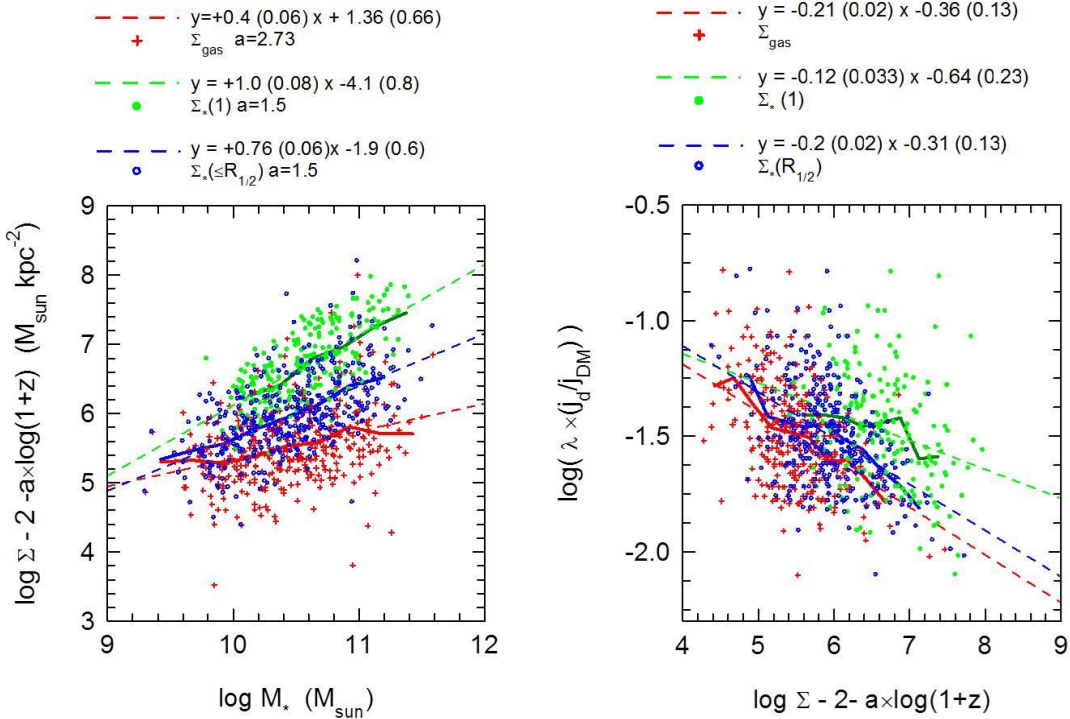

**Figure 8.** Left: $\Sigma_*(R_{1/2})$ (blue circles), $\Sigma_*(1\text{kpc})$ (filled green circles) and $\Sigma_{gas}(R_{1/2})$ (red crosses) as a function of stellar mass, after removal of their average redshift dependence (~ $(1+z)^{1.5}$ for the stellar densities (van der Wel et al. 2014a), and ~ $(1+z)^{2.7}$ for the gas surface densities (Genzel et al. 2015) ). Thick lines show the binned trend-lines, and dotted lines mark the best fit, power laws. Right: $\lambda \times (j_d/j_{DM})$ as a function of the same surface densities (identical symbols as in the left panel).

# 4. Conclusions

We have presented in this study Hα integral field unit kinematics for ~360 massive ($\log(M_*/M_\odot)$ = 9.3-11.8) $z$ = 0.8-2.6 rotationally supported disk galaxies on the star formation 'main sequence'. Our main findings are as follows.

- From the observed baryonic angular momenta of our SFGs we infer that the angular momentum parameter follows a log-normal distribution (dispersion of 0.2 in the log) around $<\lambda\times(j_d/j_{DM})>$ = 0.037 (±0.015). This distribution and its centroid do not depend on redshift, stellar or halo mass. The similarity of this angular momentum parameter is in excellent agreement with recent numerical simulations (e.g., Teklu et al. 2015; Danovich et al. 2015). Our result also lends empirical support to many theoretical models over the past three decades that assumed the disk specific angular momentum to be similar to the surrounding dark halo. Our findings are in good agreement with previous analyses of disk-dominated galaxies in the local Universe.
- There is a very significant negative correlation between $\lambda\times(j_d/j_{DM})$ and the galaxy-wide stellar and gas surface densities, but little correlation with the stellar surface densities in the central 1 kpc (tracing the bulge component). In our view this supports the proposal of Barro et al. (2013, **2015, 2016**) and Dekel & Burkert (2014) that there must be an disk-internal redistribution of angular momentum ('compaction') helping to build up massive, central bulges at $z$ ~ 1-2.5; how this redistribution affects the disk half mass radius is an interesting question that should be explored further.
- Several lines of evidence discussed in this paper and in two upcoming papers (S. Wuyts et al. 2016; P. Lang et al. 2016, in preparation) indicate that the star forming disks at the peak of the cosmic star formation activity are strongly baryon dominated, and that the mass ratio of disk to halo $m_d$ is about 5% at $\log(M_{DM}/M_\odot)$ ~ 11.1-13.3, corresponding to ~30% of the available baryons. Our results are in good agreement with recent estimates from abundance matching.

*Acknowledgements: We thank the staff of Paranal Observatory for their support. We thank Frank van den Bosch and Pieter van Dokkum for valuable comments, and Ben Moster for communicating additional information on his abundance matching analysis. We also thank the referee for a careful reading and useful suggestions to improve the paper. A.B. thanks the Astronomy Department of the University of California, Santa Cruz for their hospitality. M.F. and D.W. acknowledge the support of the Deutsche Forschungsgemeinschaft (DFG) via Project WI 3871/1-1.*

# Appendix A

## *A.1. Details of the Galaxy Sample*

The SFGs used in our analysis were taken from the following near-IR IFU samples (numbers denote disks in the 'full' sample, while numbers in brackets are for those in the 'best' sample, as described in Section 2.1):

1. 46 (26) $z = 1.5$-$2.6$ SFGs from the SINS (Förster Schreiber et al. 2009) and zCOSMOS (zC)-SINF surveys (Mancini et al. 2011; N. M. Förster Schreiber et al. 2016, in preparation), of which 33 (14) were observed with adaptive optics (AO) ($R_{1/2,\mathrm{beam}}$ = FWHM/2 ~ 0.1″), while the rest were observed in seeing limited mode ($R_{1/2,\mathrm{beam}}$ ~ 0.25″-0.3″) using SINFONI on the ESO VLT (Eisenhauer et al. 2005; Bonnet al. 2006).

2. 273 (206) $z = 0.76$-$2.6$ SFGs from the ongoing KMOS$^{3D}$ survey (Wisnioski et al. 2015), all observed in seeing limited mode ($R_{1/2,\mathrm{beam}}$ ~ 0.2″-0.35″) with the KMOS multiplexed IFU instrument on the VLT (Sharples et al. 2008; 2012). This sample represents the results of the first two years of the five-year KMOS$^{3D}$ survey.

3. 7 (0) $z = 1.3$-$2.6$ SFGs are from the AO-assisted IFU data sets of Law et al. (2009) and Wright et al. (2007, 2011), observed with OSIRIS at the Keck telescope (Larkin et al. 2006; Wizinowich et al. 2006).

4. 25 (0) $z = 0.9$-$1.5$ SFGs from the MASSIV survey (Épinat et al. 2009, 2012; Contini et al. 2012), 23 (0) observed in seeing limited mode, and 2 (0) observed in AO mode with SINFONI.

5. 6 (0) $z = 0.8$-$1.46$ SFGs from the HiZELS SINFONI sample of Swinbank et al. (2012), all observed in AO-assisted mode.

6. In addition we also included 1 (1) $z = 1.6$ SFG observed in seeing limited, slit scanning mode with the LUCI slit spectrometer on the Large Binocular Telescope (Genzel et al. 2013), and 1 (0) $z = 2$ SFG from the FIRES survey observed in seeing limited mode with SINFONI (van Starkenburg et al. 2008).

In the rest of this Appendix, we summarize the derivation of the global stellar properties, and of the structural and kinematic parameters of the galaxies from our SINS/zC-SINF, KMOS$^{3D}$, and LUCI data sets, which form the vast majority of the disk sample studied in this paper (320 out of the 359 of the 'full' sample, and all 233 of the 'best' sample). For galaxies from the other IFU samples, we adopted the properties as reported in the respective papers listed above whenever they are available and derived consistently with our procedures (with adjustments where necessary, for example to scale the stellar masses and SFRs to our adopted Chabrier (2003) IMF) or we derived them based on published data following our methodology.

## A.2. Stellar Properties, Structural and Kinematic Analysis, and Beam Smearing Corrections

### A.2.1. Stellar Masses, Star Formation Rates, and Gas Masses

The global stellar properties were derived following the procedures outlined by Wuyts et al. (2011a). In brief, stellar masses were obtained from fitting the observed broadband optical to near-/mid-IR (rest-UV to optical/near-IR) spectral energy distributions (SEDs) with Bruzual & Charlot (2003) population synthesis models, adopting the Calzetti et al. (2000) reddening law, the Chabrier (2003) IMF, a solar metallicity, and a range of star formation histories (in particular including constant SFR, as well as exponentially declining or increasing SFRs with varying e-folding timescales). Of the parameters fitted in the modeling (which include stellar mass and age, visual extinction, and star formation history), the stellar mass tends to be the most robust parameter especially for SEDs that extend to the rest-frame near-IR, as is the case for most of the SINS-zC-SINF, KMOS$^{3D}$, and LUCI SFGs (e.g., Papovich et al. 2001; Förster Schreiber et al. 2004, 2009; Shapley et al. 2005; Wuyts et al. 2007, 2011a; Maraston et al. 2010). Over the mass and redshift ranges of the galaxies, gas-phase O/H abundances inferred from rest-optical nebular emission lines suggest metallicities of ~ 1/4 to ~ 1× solar (E. Wuyts et al. 2014, 2016; see also, e.g., Erb et al. 2006; Zahid et al. 2011, 2014; Stott et al. 2013; Steidel et al. 2014; Sanders et al. 2015). Varying the assumed metallicity in this range would change the stellar masses in our modeling by < 0.1 dex (e.g., Wuyts et al. 2007; Förster Schreiber et al. 2009). Given the uncertainties in metallicity determinations for high-$z$ SFGs (see, e.g., Kewley et al. 2013 and references therein), known degeneracies with other model parameters in broadband SED modeling, and the small impact on derived stellar masses, we chose to keep a fixed solar metallicity. We note that throughout the paper, we define stellar mass as the 'observed' mass ('live' stars plus remnants), after mass loss from stars. This is about 0.15 to 0.2 dex smaller than the integral of the star formation rate over time.

The SFRs were obtained from rest-frame UV + infrared luminosities through the Herschel-Spitzer-calibrated ladder of SFR indicators of Wuyts et al. (2011a) or, if not available, from the broadband SED modeling described above.

Individual determinations of molecular gas masses (from CO line or submillimeter/far-infrared dust continuum emission) are available only for a very small number of our galaxy sample, and atomic hydrogen masses are not known for any of our high-$z$ SFGs. Instead, we computed molecular gas masses from the general scaling relations between star formation rates, stellar masses, and molecular gas masses for main sequence SFGs (as a function of redshift) as presented by Genzel et al. (2015). We assumed, as argued in that paper, that at $z \sim$ 1-3 the cold gas content of SFGs is dominated by the molecular component such that the atomic fraction can be neglected. As such the gas masses estimated from these scaling relations may be lower limits.

For the main-sequence SFG population (with near constant star formation histories), we adopted uncertainties of ±0.15 dex for the stellar masses, and ±0.2 dex for the star formation rates, although somewhat smaller uncertainties may be appropriate for SFGs with measurements of individual far-infrared luminosities (Wuyts et al. 2011a). For the gas masses, we adopted uncertainties of ±0.2 dex (Genzel et al. 2015).

### A.2.2. Kinematic Parameters and Classification

As mentioned in the Introduction, recent work has established that the strong majority of main-sequence, star forming galaxies at $z \sim$ 0.8-2.6 are turbulent (thick), rotating disks with approximately exponential stellar light/mass profiles. In our data analysis we extracted the Hα velocity field by fitting Gaussian line profiles to each IFU spatial pixel, in some cases after some prior smoothing to increase signal to noise ratios, resulting in spatially resolved maps of the velocity centroids and velocity dispersions from which we derived the kinematic parameters of interest, $v_{rot}$ and $\sigma_0$. The quantity $v_{rot}$ is the maximum rotational velocity corrected for beam smearing and inclination $i$ ($v_{rot} = c_{psf,v} \times v_{obs}/\sin i$), and $\sigma_0$ is the intrinsic velocity dispersion corrected for beam smearing ($\sigma_0 = c_{psf,\sigma} \times \sigma_{obs}$). Here $v_{obs}$ is half of the difference between the maximum positive and negative velocities on both sides of the galaxy, $\sigma_{obs}$ is the measured line width in the outer parts of the galaxy *corrected for instrumental spectral resolution* (i.e., subtracting in quadrature $\sigma_{instr}$), and $c_{psf,v}$ and $c_{psf,\sigma}$ are beam smearing corrections for the velocity and velocity dispersion, respectively. The median ratio of the intrinsic half-light radius of the galaxies to the radius of the point spread function associated with their data set, $b = R_{1/2}/R_{1/2,beam}$, is 1.7 for the SFGs in the 'full' sample, and about 12% of that sample have a $b < 1$. This means that beam smearing is significant, and lowers the amplitude of maximum velocity gradient and increases the intrinsic velocity dispersion. Sections A.2.3 and A.2.4 below describe how the galaxies' radii, inclinations, and beam smearing corrections were derived.

Following Wisnioski et al. (2015), we classified a galaxy as a 'rotation dominated' disk if

1. the velocity map exhibits a continuous velocity gradient along a single axis. In larger systems with good signal to noise ratio this is synonymous with the detection of a 'spider' diagram in the two-dimensional, first moment velocity map (van der Kruit & Allen 1978);

2. $v_{rot}/\sigma_0 > 1.5$-$2$; given instrumental uncertainties we use $v_{rot}/\sigma_0 = 1.5$ and 2 to distinguish 'rotation dominated' from 'dispersion dominated galaxies' in the 'full' and 'best' samples, respectively;

3. the position of the steepest velocity gradient, as defined by the midpoint between the velocity extrema along the kinematic axis, is coincident within the uncertainties with the peak of the velocity dispersion map;

4. the photometric and kinematic axes are in agreement ($\leq 30$ degrees); and

5. the kinematic center of the galaxy coincides with the maximum/centroid of the stellar distribution.

As discussed by Wisnioski et al. (2015) for the seeing limited KMOS$^{3D}$ survey, 83% of the resolved galaxies fulfill criteria 1 and 2 (92% at $z \sim 1$ and 74% at $z \sim 2$). This fraction slowly drops if the stricter criteria 3, 4 and 5 are added, and amounts to 70% if all 5 criteria are used. Similar results are obtained in the other recent surveys, or if higher resolution AO data sets are considered (e.g., Newman et al. 2013; Genzel et al. 2014; Tacchella et al. 2015a; N. M. Förster Schreiber et al. 2016, in preparation, for the SINS/zC-SINF sample).

*A.2.3. Inclinations and Disk Radii*

With the exception of the most massive SFGs, the stellar surface brightness distributions of main-sequence SFGs across the mass- and redshift range discussed in this paper are reasonably well fit by near-exponential (Sérsic index $n_S$~1-1.5) profiles (Wuyts et al. 2011b; Bell et al. 2012; Bruce et al. 2014a,b; Lang et al. 2014). For this reason, our starting assumption is that stars and gas in all rotation-dominated SFGs of our 'full' and 'best' samples are distributed in symmetric oblate, thick disks with the same exponential profile (for corrections to variable Sérsic indices, see Appendix B.7). Based on the statistical distribution of projected minor to major axis ratios in the $z = 0.5$-3 3D-HST/CANDELS reference sample, this assumption is quite well justified for the massive ($\log(M_*/M_\odot)>10$) SFG population constituting the large majority of our sample. The justification appears to break down at lower masses, where triaxial systems become common (Law et al. 2012; van der Wel et al. 2014b). These triaxial systems are plausibly identical to the dispersion dominated galaxies that we have eliminated from our sample. For symmetric oblate disks, inclinations can be determined from the morphological minor to major axis ratio, $b/a$, such that $\cos^2(i) = ((b/a)^2 - \kappa^2)/(1-\kappa^2)$, with $\kappa$ ~0.15-0.2 at $z$ ~ 1-3 (Law et al. 2009; Förster Schreiber et al. 2009; Wisnioski et al. 2015).

For all of the KMOS$^{3D}$ (and also MASSIV), and most of the SINS/zC-SINF galaxies, inclinations $i$ and half-light (effective) radii $R_{1/2}$ were inferred from Sérsic model fits to the rest-frame optical stellar light distributions available from broadband imaging with HST (or from the ground for MASSIV). For the remainder of the SINS/zC-SINF galaxies (and for the OSIRIS and HiZELs samples), half-light radii were inferred from the line integrated Hα distributions while the inclinations were inferred from the continuum images synthesized from the IFU data. To first order this approach is justified as high-$z$ SFGs are gas-rich with large star formation rates and young stellar populations. However, the presence of substantial stellar bulges in the more massive high-$z$ SFGs (e.g., Lang et al. 2014), with lower Hα equivalent widths than in the disks, results in the ionized gas disks being somewhat more extended than the stellar distributions (e.g., Genzel et al. 2014; Tacchella et al. 2015b). This has been compellingly demonstrated in a recent comparison of the rest-frame *R*-band continuum and Hα emission sizes in the 3D-HST survey. From Hα image stacking of 2000 $0.7 < z < 1.5$ SFGs, Nelson et al. (2015) found that the average ratio of Hα to continuum size is <$R_{1/2}$(Hα)/$R_{1/2}$(*R*-band)> =1.1 × ($M_*/10^{10} M_\odot)^{0.05}$ (see also Förster Schreiber et al. 2011a; Nelson et al. 2012; Wuyts et al. 2013).

### *A.2.4. Beam Smearing Corrections*

To infer the intrinsic maximum disk rotation velocity near the half-light radius (~1.2 $R_{1/2}$ for a well resolved thin exponential disk, neglecting dark matter), one needs to correct for the effect of beam smearing, either by fitting each data cube with a disk model or, alternatively, by employing scaling relations from observed to intrinsic rotation velocity. We have used the former approach in several of our recent papers, especially when analyzing high resolution, adaptive optics data sets and trying to establish full rotation curves (Genzel et al. 2006, 2008, 2011, 2014; Cresci et al. 2009; S. Wuyts et al. 2016). For the analysis in this paper we use the second approach, since we are mainly interested in extracting the value of the maximum rotation velocity, and since most of our data sets are in seeing limited mode. For galaxies with a reliable disk model in the sample studied here, the rotation velocities derived from both approaches agree well, to better than 10% on average and within the uncertainties.

Assuming exponential mass distributions as motivated in the last section we computed mock data cubes as a function of stellar mass, inclination, disk exponential scale length, intrinsic velocity dispersion and instrumental resolution using DYSMAL (Davies et al. 2011), which creates 'observed' data cubes by convolving the intrinsic cubes with the instrumental beam spectrally and spatially. For seeing limited cubes we used a Gaussian PSF kernel of the appropriate FWHM, while for SINFONI AO data sets we used a double Gaussian PSF kernel to reflect the combination of the diffraction limited core and residual seeing on the beam profile. The ratio between the maximum intrinsic rotation velocity of an exponential distribution at ~2 $R_d$ (~ 1.2 $R_{1/2}$) to the observed rotation velocity, which we will call the velocity beam correction factor, $c_{psf,v}$, is very well described by a double parameter function, which depends on the ratio $x = R_{1/2}/R_{1/2,beam}$, as well as on the ratio of the radius $R_{vel}$ at which the observed velocity gradient was determined and the half-light radius, $y = R_{vel}/R_{1/2}$. Variations in all other parameters introduce only secondary changes that are negligible. Tables A1 and A2 give fitting functions $c_{psf,v}(x,y)$ for single-Gaussian (seeing limited) and double-Gaussian (SINFONI AO) point spread functions. As an example the left panel in Figure 9 shows these fitting functions for the single- and double- Gaussian kernels for $R_{vel}/R_{1/2} = 1$. As expected the correction factors become large if $R_{1/2}/R_{1/2,beam}$ is below one. In that case $c_{psf,v}$ becomes very sensitive to small deviations from the assumed exponential distribution, for instance because of a central bulge, or because there are bright star forming clumps outside the nuclear region. For these reasons, we decided to include in the final analysis only those SFGs with $c_{psf,v} <$ 5 for the 'full' and $c_{psf,v} < 2$ for the 'best' samples. However, we find that none of the results reported in this paper depend on this choice, indicating that the beam smearing corrections are robust.

Another important parameter for our analysis is the intrinsic velocity dispersion, or alternatively the vertical scale height of the disk. Analysis of the best AO IFU data sets currently available indicates that this intrinsic velocity dispersion is constant to first order within a galaxy (Genzel et al. 2011; N. M. Förster Schreiber et al. 2016, in preparation). However, we caution that the still modest resolution of current IFU AO data when compared to the angular sizes of high-$z$ galaxies, combined with the impact of beam-smeared large scale streaming motions (e.g., rotation) appearing as an increased velocity dispersion, make detailed statements on the spatial variations of the velocity dispersion difficult, especially near the kinematic center. The constant velocity dispersion floor $\sigma_0$ does appear to vary modestly from galaxy to galaxy at a given redshift. Most importantly, $\sigma_0$ decreases with decreasing redshift ($\sigma_0 \sim$ 18×(1+$z$) km s$^{-1}$; Wisnioski et al. 2015; see also Kassin et al. 2012).

**Table A1. Velocity beam smearing corrections for a Gaussian PSF**

$c_{psf,v} = v_{rot}(R=R_{1/2})|_{intrinsic}/(v_{obs}(R=R_{vel}) \times \sin^{-1}(i))$ as a function of $x = R_{1/2}/R_{1/2,beam}$ and $y=R_{vel}/R_{1/2}$: $c_{psf,v}(x,y) = 1 + A(y) \times \{B(y) + x\}^{C(y)}$, where $R_{1/2,beam}$ is the PSF HWHM.

| y | A | B | C |
|---|---|---|---|
| 1 | 1.28 | -0.15 | -1.78 |
| 1.5 | 0.58 | -0.25 | -1.60 |
| 2 | 0.34 | -0.44 | -0.86 |
| 2.5 | 0.30 | -0.50 | -0.40 |

**Table A2. Velocity beam smearing corrections for a double Gaussian**

$c_{psf,v} = v_{rot}(R=R_{1/2})|_{intrinsic}/(v_{obs}(R=R_{vel}) \times \sin^{-1}(i))$ as a function of $x = R_{1/2}/R_{1/2,beam}$ and $y=R_{vel}/R_{1/2}$: $c_{psf,v}(x,y) = 1 + A(y) \times \{B(y) + x\}^{C(y)}$, where $R_{1/2,beam}$ is the HWHM of the AO PSF core component (0.08″ for our SINFONI AO data).

| y | A | B | C |
|---|---|---|---|
| 1 | 1.56 | -0.30 | -1.10 |
| 1.5 | 1.15 | -0.27 | -1.15 |
| 2 | 1.25 | -0.13 | -1.29 |
| 2.5 | 1.96 | 0.15 | -1.60 |

Again we used our simulated data sets to determine correction factors between the measured velocity dispersion extracted in the outer disk parts ($R$~2-2.5 $R_{1/2}$), where the influence of beam smeared rotation is minimal (with the instrumental spectral resolution already removed). The right panel of Figure 9 depicts these dispersion correction factors. In this case the correction factors depend on inclination, stellar mass, and intrinsic dispersion, such that we created look-up tables to then estimate the beam corrected intrinsic velocity dispersion, which for the purpose of this analysis we assumed to be spatially constant across the galaxy (see however the discussion in Appendix B.3. below).

After correction, our 'full' sample of 359 SFGs consists of 334 rotation-dominated disks, for which the inclination and beam smearing corrected ratio of the rotation velocity at the peak of the rotation curve ($v_{rot}=v_{rot}(R\sim R_{1/2})$) to the local velocity dispersion in the outer disk $\sigma_0 \sim \sigma_0(R\sim 1.5\text{-}2.5\ R_{1/2})$ is $v_{rot}/\sigma_0 \geq 2$. This includes 4 objects which may be in the process of a minor merger but for which the rotation curve of the main galaxy does not appear to be significantly disturbed. Another 23 SFGs have $1.5 \leq v_{rot}/\sigma_0 < 2$, which could be either rotating disks or dispersion-dominated, given the typical uncertainties of $\Delta(v_{rot}/\sigma_0) \sim 0.5$. Two of these may be a minor merger. All of our 'best' SFGs (233 galaxies) are well resolved rotating disks, with one object having a very small neighbor that does not seem to affect the kinematics of the main disk.

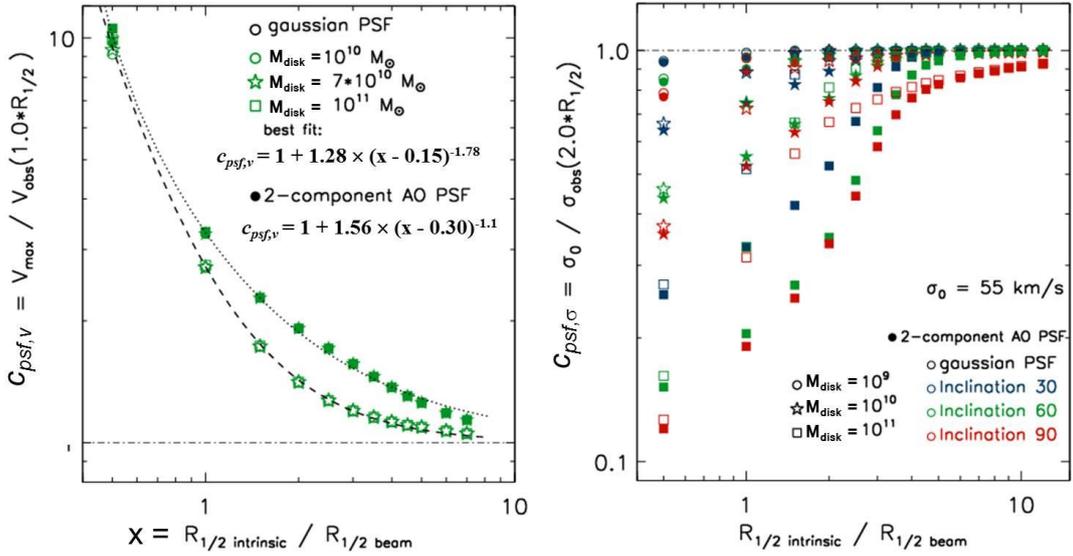

**Figure 9.** Left panel: Correction factor ($c_{psf,v}$) between the observed maximum rotation velocity and the intrinsic maximum rotation velocity of an exponential disk mass distribution at $R \sim R_{1/2}$ (1.68 $R_d$) for a single Gaussian point spread function (open symbols, appropriate for seeing limited observations, such as in KMOS$^{3D}$), as well as for a double Gaussian point spread function appropriate for AO observations (filled symbols). Circles, stars, and squares show the corrections for different disk masses as labeled in the plot. The results from the mock data sets can be well described by a fitting function that depends on the ratio of $R_{1/2}$ to beam size $R_{1/2,beam}$ and the ratio of the radius at which the velocity was determined, $R_{vel}$, relative to $R_{1/2}$ (in the graph we show the case $R_{vel}=R_{1/2}$). Right panel: Correction factor from the observed velocity dispersion at $R \sim 2 \times R_{1/2}$ to the intrinsic velocity dispersion (assumed in this specific simulation to be $\sigma_0 = 55$ km/s, appropriate for $z \sim 2\text{-}2.6$, Wisnioski et al. 2015), again as a function of $R_{1/2}/R_{1/2,beam}$ for different stellar masses and inclinations (symbol shapes and colors, respectively, as labeled in the plot), and again for seeing limited (open symbols) and AO observations (filled symbols). The outer velocity dispersion correction factor cannot be described by a single parameter fitting function and a number of such graphs (for different $\sigma_0(z)$) have to be used to create a look-up table to estimate the intrinsic velocity dispersion for each galaxy.

# Appendix B

## *B.1. Isothermal Disk Model*

The simplest assumption that one can make is a ***completely dark matter dominated disk***. In this case its rotation velocity directly traces the dark halo mass distribution and by this also its virial parameters. Mo et al. (1998; see also the earlier work by Fall & Efstathiou 1980 and Fall 1983) derived simple expressions for λ and $M_{DM}$, adopting a non-self-gravitating, exponential disk, embedded in an isothermal halo with a truncation radius at $R_{virial}$. Combination of equations (1) and (3) yields

$$\lambda = \sqrt{2} \times \left(\frac{R_d}{R_{virial}}\right) \times \left(\frac{j_d}{j_{DM}}\right)^{-1}$$

$$\stackrel{\text{DM dominated isothermal disk}}{=} 10 \cdot \sqrt{2} \times H(z) \times \left(\frac{R_d}{v_{rot}}\right) \times \left(\frac{j_d}{j_{DM}}\right)^{-1} \qquad (B1).$$

λ increases linearly with disk scale length $R_d=R_{1/2}/1.68$, as expected. However, it decreases with increasing $v_{rot}$, which is somewhat counter-intuitive as for a given radius angular momentum scales linearly with rotational velocity. This anti-correlation results from the fact that λ is defined as the ratio of the specific angular momentum of the disk, which is proportional to $R_d \times v_{rot}$, divided by the product of $R_{virial} \times v_{virial} \sim v_{virial}^2$. For a completely dark matter-dominated disk, $v_{rot} \sim v_{virial}$, which leads to λ ~ $R_d/v_{rot}$.

This '***isothermal model***' in the bottom line of equation (B1) has serious caveats. First of all, the self-gravity of galactic disks is not negligible. For most high-redshift galaxies, the (baryonic) disk mass dominates the rotation curve inside $R_{1/2}$ (see section 3.3; Förster Schreiber et al. 2009; S. Wuyts et al. 2016). Second, dark matter halos are not isothermal (Navarro et al. 1996, 1997 (NFW)). Finally, the velocity dispersion $\sigma_0$ has been neglected, which can significantly affect rotation curves in the outer regions, especially for galaxies with small values of $v_{rot}/\sigma_0$ (Burkert et al. 2010).

We will now discuss how we included these aspects in our modeling.

## B.2. Exponential Disk within an NFW Dark Matter Halo

Following Mo et al. (1998), we now focus on a second approach, assuming that an exponential baryonic disk with surface density distribution $\Sigma_d(R) = \Sigma_0 \exp(-R/R_d)$ is embedded in an NFW dark matter halo. Its circular velocity is given by the sum of the disk and halo contribution

$$v_{circ}^2 = v_{disk}^2 + v_{DM}^2 \qquad (B2),$$

with the 'thin disk limit' (Freeman 1970; Navarro et al. 1997)

$$v_{disk}^2(y = R/2R_d) = 4\pi G \Sigma_0 R_d y^2 \times \left[ I_0(y) K_0(y) - I_1(y) K_1(y) \right],$$

and

$$v_{DM}^2 = v_{virial}^2 \times \left( \frac{R_{virial}}{R} \right) \times \frac{\ln(1 + R/R_s) - (R/R_s)(1 + R/R_s)}{\ln(1+c) - c/(1+c)} \qquad (B3),$$

where $I_n$ and $K_n$ denote modified Bessel functions of order $n$, $c$ is the concentration parameter of the halo (Bullock et al. 2001b), and $R_s = R_{virial}/c$ (NFW). For thick disks, as appropriate at high-$z$, the disk rotation velocity at $\sim R_{1/2}$ is about 10% greater than given in equation (B3) (Noordermeer 2008).

Mo et al. (1998) assumed that once a self-gravitating baryonic disk forms, the dark halo contracts adiabatically. However, feedback from supernovae, massive stars, and AGN act to expand the halo. Burkert et al. (2010) found from their analysis that fitting the observed kinematics of high-$z$ disks including adiabatically contracted dark halos would require extreme baryon fractions that could even exceed the cosmic baryon fraction. They therefore concluded that dark halos did not contract substantially during gas infall and disk formation. ***We thus took as our default a model without adiabatic halo contraction***. The sample of galaxies studied in detail by Burkert et al. (2010) was small and therefore the conclusions of no significant adiabatic contraction might not apply to all galaxies studied here. In a second step, we therefore also investigated models with adiabatic contraction and found that the results do not change much, other than in slightly increased angular momentum parameters and disk to dark halo mass fractions. More detailed studies of galaxy rotation curves are required to settle the question of adiabatic contraction (P. Lang et al. 2016, in preparation).

## B.3. Disk Truncation due to Turbulent Pressure

In hydrostatic equilibrium a turbulent disk with one-dimensional velocity dispersion $\sigma$ and mid-plane density $\rho$ has a scale height $h = \sigma/\sqrt{2\pi G\rho}$. The observed rotational velocity $v_{rot}$ deviates from $v_{circ}$ as the turbulent pressure gradient $d(\rho\sigma^2)/dR$ leads to an additional radial force ('asymmetric drift'), requiring the centrifugal force and by this $v_{rot}$ to be adjusted in order to match the gravitational force. Here we neglect thermal pressure gradients, as the thermal sound speed is in general small compared to the turbulent velocity. The reduced rotational velocity as a function of radius is given by (Binney & Tremaine 2008; Burkert et al. 2010)

$$v_{rot}^2 = v_{circ}^2 + 2\sigma^2 \times \frac{d\ln\Sigma}{d\ln R} = v_{circ}^2 - 2\sigma^2 \times \left(\frac{R}{R_d}\right) \quad (B4).$$

Equation (B4) is valid even if $\sigma$ is a function of $R$. Note that if the surface density distribution is not exponential, but a more general Sérsic distribution of index $n_S$ ($\Sigma = \Sigma_0 \exp(-b_{n_S}\left(R/R_{1/2}\right)^{n_S})$), then the last term on the right side of equation (B4) becomes $-2\sigma^2 \times b_{n_S} \times (R/R_{1/2})^{1/n_S}$. For high dispersions the rotational velocity can be strongly reduced in the outer disk regions, leading to a decline in rotation that could be even steeper than Keplerian ($v_{rot}^2 \sim R^{-1}$).

The observations provide an estimate of $\sigma = \sigma_0$ at $\sim$ 2-2.5 $R_{1/2}$, which we adopted as the characteristic dispersion everywhere in the disk (see Appendix A.2). According to equation (B4), this isothermal disk has a finite 'truncation' radius $R_{max}/R_d = 0.5 \times (v_{circ}/\sigma_0)^2 \sim$ 2-15 where $v_{rot} = 0$. The total cumulative mass of an exponential disk within a given radius is

$$M(<R) = 2\pi\Sigma_0 R_d^2 \times \left[1 - \left(1 + \frac{R}{R_d}\right) \times exp\left(-\frac{R}{R_d}\right)\right] \quad (B5).$$

Its half-mass radius is defined as $M_d(<R_{1/2}) = 0.5 \times M_d(<R_{max})$, which with equation (B5) leads to an implicit equation for $R_{1/2}/R_d$:

$$1 - \left(1 + \frac{R_{1/2}}{R_d}\right) \times \exp\left(-\frac{R_{1/2}}{R_d}\right) =$$

$$0.5 \times \left[1 - \left(1 + 0.5\left(\frac{v_{circ}}{\sigma_0}\right)^2\right) \times \exp\left(-0.5\left(\frac{v_{circ}}{\sigma_0}\right)^2\right)\right] \quad \text{(B6)}.$$

The solid and dashed lines in Figure 10 show that $R_{1/2}/R_d$ is uniquely specified by $v_{rot}/\sigma_0$ or $v_{circ}/\sigma_0$, measured at the half-mass radius. Remember that $v_{rot}$ is the observed rotational velocity while $v_{circ}$ represents the rotation velocity of a disk with negligible dispersion. For kinematically cold disks with large ratios of rotation-to-dispersion, $v_{rot} = v_{circ}$ and the solution approaches the constant value $R_{1/2} = 1.68 \times R_d$. Disks with larger velocity dispersions can however be strongly dispersion truncated with half-mass radii that can become even smaller than $R_d$. A convenient approximation that is shown by the filled points in Figure 10 is

$$\frac{R_{1/2}}{R_d} = 1.7 \left(\frac{v_{1/2}}{\sigma_0}\right)^{2.7} \times \left(1 + \left(\frac{v_{1/2}}{\sigma_0}\right)^{2.7}\right)^{-1} \quad \text{(B7)}.$$

Here, $v_{1/2} = v_{rot}(R_{1/2})$.

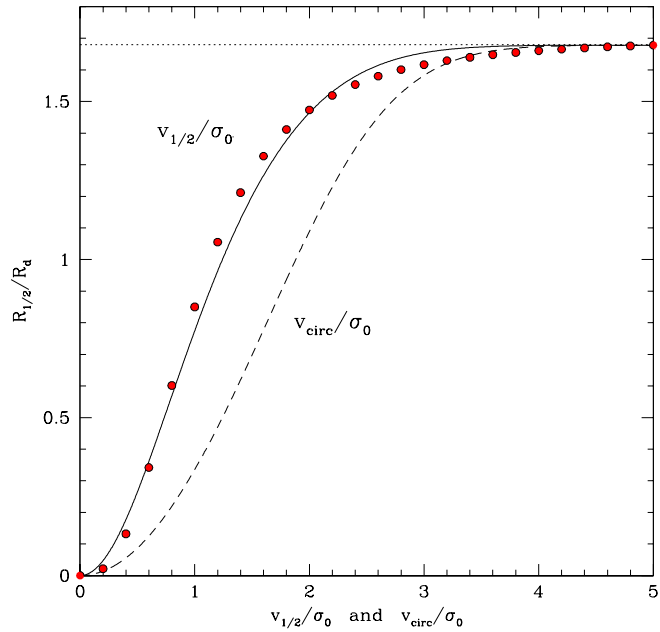

**Figure 10.** The ratio of disk half mass radius to exponential scale radius as a function of $v_{1/2}/\sigma_0$ and $v_{circ}/\sigma_0$. The red points show the approximation given by equation (B7). The dotted line shows the asymptotic limit of 1.68 for strongly rotationally supported disks.

## B.4. Determination of $M_{DM}$ and $\lambda$

Given the observed redshift, the disk's half-mass radius, baryonic disk mass, rotational velocity $v_{1/2}$ and velocity dispersion $\sigma_0$, equations (B1) to (B7) in principle uniquely specify the dark matter halo's parameters $M_{DM}$ (and thus $R_{virial}$ from equation (1)), since $v_{1/2}$ contains contributions from both disk and halo (equation (B2)), and since the disk mass is assumed to be known from the sum of stellar and (molecular) gas mass. Equation (B1) then yields $\lambda$, the angular momentum parameter of the dark matter halo. More precisely, the knowledge of $R_d$ and $R_{virial}$ ($M_{DM}$) yields $\lambda \times (j_d/j_{DM})$ from equation (3). The disk's total and specific angular momenta are directly determined by integrating $J_d = 2\pi \int_0^{R_{max}} \Sigma_d v_{rot} R^2 dR$, and the total disk mass, which is an integral over the disk surface density.

## B.5. Monte-Carlo modeling

In practice, the combination of observational uncertainties, the high baryonic fraction within $\sim R_{1/2}$ (see Section 3.3) and the uncertainties in the theoretical assumptions (e.g., adiabatic contraction) make the individual estimates of $M_{DM}$ quite uncertain. In order to evaluate how observational uncertainties affect the results we performed a Monte-Carlo study, adopting $N = 10{,}000$ randomly chosen values of ($R_{1/2}$, $v_{1/2}$, $\sigma_0$, $\log(M_d)$) centered around the observed values with a Gaussian probability with half-width half maximum as given by the observational uncertainties. Not all combinations of parameters lead to a reasonable model. We discarded all solutions with total disk baryon fractions $m_d$ larger than 25%, in order not to violate the cosmic baryon fraction. The range of allowed solutions then specifies the average value and error in $M_{DM}$ and $\lambda \times (j_d/j_{DM})$. As an example, Figure 11 shows the result of the Monte-Carlo simulation for BX 455. In order to suppress overcrowding only 1000 points are shown. Blue points in the upper two panels show the distribution of disk parameters, centered on the observed values (large red triangles) that lead to a theoretically converged model. The systematic offset between the red and blue points results from the fact that certain systematic combinations of the model parameters lead to values of $m_d$ that violate the cosmic baryon fraction and therefore are discarded. The large cyan circle shows the mean of all converged models. The blue points in the lower left panel show the corresponding dark halo mass and the spin parameter of the disk. The cyan circle shows the mean values with uncertainties. For BX 455 we infer a dark matter-to-baryon mass fraction of $M_{DM}/M_{baryon}$ ~19.5 and a lambda parameter of log $\lambda$ = -1.55±0.22. As expected from the parameter dependences, the uncertainties in individual dark matter masses is substantial (±0.35 dex), but the typical uncertainty in $\lambda$ is lower (±0.21 dex).

At the end of this exercise, we obtained good converged fits for $M_{DM}$, $M_{DM}/M_d$ and $\lambda$ for 321 of the 359 SFGs in the 'full' sample, and 220 of the 233 SFGs in the 'best' disk sample.

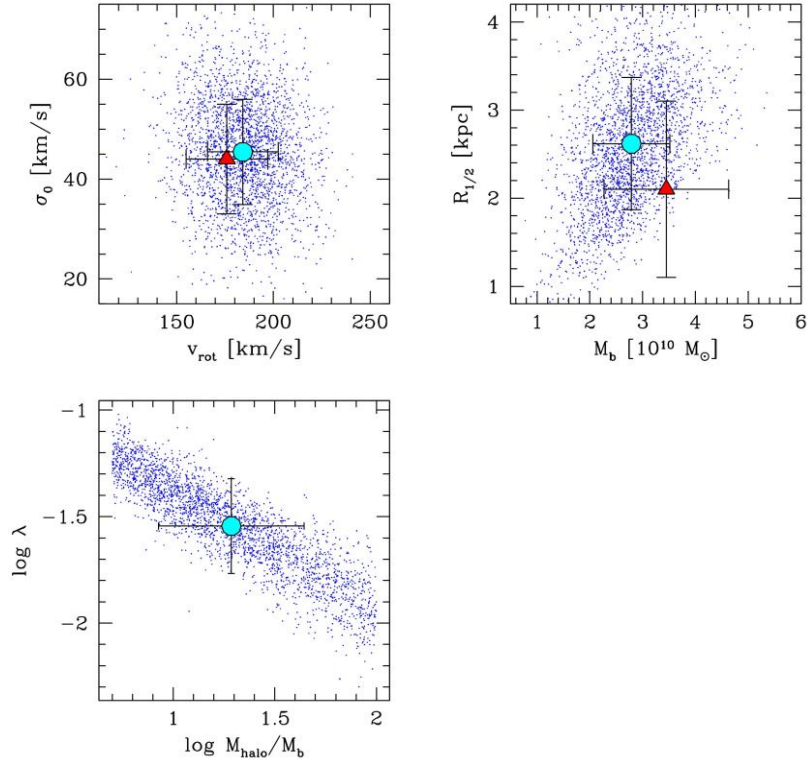

**Figure 11.** NFW Monte-Carlo modeling of BX 455. The large red triangles with error bars show the galaxy's observed physical properties and uncertainties. Small blue points correspond to converged disk-halo models, drawn randomly with mean values and standard deviations as given by the observations. The large cyan circles with error bars depict the mean values and standard deviations of these data points. The upper left diagram shows the velocity dispersion versus the disk rotation at $R_{1/2}$, the upper right panel shows the half mass radius versus the disk's total baryonic mass. The lower left diagram depicts the corresponding dark matter mass fraction and disk lambda parameter, respectively, as well as their mean values (cyan circle) and standard deviation.

## B.6. Determination of λ from adopted $m_d$-relations

Another estimate of the angular momentum parameter of the halo can be obtained from the observed disk parameters if the ratio of disk to halo mass is known. For instance, one may assume that $m_d$ has the same constant value for all SFGs. Alternatively, one can invert the relations $M_*/M_{DM}$ versus $M_{DM}$ obtained from abundance matching (Moster et al. 2010, 2013; Behroozi et al. 2010, 2013a) to infer $M_d/M_{DM}$ versus $M_*$. In that case one can write for an NFW halo of concentration $c$ (e.g., Romanowsky & Fall 2012)

$$j_{DM} = q_{DM}(c) \times \lambda \times R_{virial} \times v_{virial} \quad \text{, and with equation (1)}$$

$$\lambda = \frac{q_d \times R_d \times v_{1/2}}{q_{DM} \times H^{-1/3}(z) \times M_{DM}^{2/3}} \times \left(\frac{j_{DM}}{j_d}\right)$$

$$\propto \frac{q_d}{q_{DM}} \times R_d \times v_{1/2} \times \left(\frac{j_{DM}}{j_d}\right) \times \left(\frac{M_d}{m_d}\right)^{-2/3} \quad (B8).$$

Here $q_d$ and $q_{DM}(c)$ are coefficients that relate the disk's and halo's product of rotational velocity and radius to the specific angular momentum (e.g., for a thin disk with $\sigma_0 = 0$, $q_d = 2$). For a turbulent exponential disk with the properties described in the last sections we find the following fitting function for determining λ

$$\lambda = f \times 5.16 \cdot 10^{-5} \times \left(\frac{j_{DM}}{j_d}\right) \times (0.0153c + 1) \times \left(\frac{R_d \times v_{1/2}}{\text{kpc} \times \text{km/s}}\right) \times \left(\frac{H(z)}{100 \text{ km/s/kpc}}\right)^{1/3} \times \left(\frac{M_d/m_d}{10^{12} M_\odot}\right)^{-2/3}$$

$$f = 1 \quad \text{for } (v_{1/2}/\sigma_0) > 10$$
$$f = 0.5\left[1 + \cos\left(\pi(\sigma_0/v_{1/2} - 0.1)\right)\right] \quad \text{for } 10 \geq (v_{1/2}/\sigma_0) \geq 1.4 \quad (B9).$$
$$f = 0.35 \quad \text{for } (v_{1/2}/\sigma_0) < 1.4$$

The fitting function $f$ encapsulates the dependence of $q_d$ on $v_{1/2}/\sigma_0$ and thus, on the truncation of the disk discussed in section B.3.

## B.7. Impact of Deviations from Exponential Distributions

So far we have assumed that the surface density distribution of the baryons is exponential ($n_S = 1$). The analysis of the *H*-band light from HST imaging of the reference 3D-HST/CANDELS galaxy population (described in Section 2.1) suggests this is roughly correct for SFG galaxies on the main sequence although there is a trend of increasing Sérsic indices above unity at $\log(M_*/M_\odot)>10.5$ (Wuyts et al. 2011b; Bell et al. 2012; Lang et al. 2014). To investigate the impact of variations in Sérsic index and dispersion truncation we followed Romanowsky and Fall (2012) and computed

$$j_d(n_S, y = \log(\sigma_0/v_{1/2})) = \int_0^{R_{max}(y)} \Sigma(R, n_S) \times v_{rot}(R, n_S, y) \times R^2 dR \Big/ \int_0^{R_{max}(y)} \Sigma(R, n_S) \times R \, dR,$$

and

$$k(n_S, y) = \frac{j_d(n_S, y)}{(v_{1/2} \times R_{1/2})}, \text{ with}$$

$$\Sigma(R, n_S) = \exp\left(-b_{n_S}\left(-\frac{R}{R_{1/2}}\right)^{1/n_S}\right) \text{ and } b_{n_S} = 2n_S - 1/3 + 0.009876/n_S \qquad \text{(B10).}$$

for a grid of points in ($x=\log(n_S)$, $y=\log(\sigma_0/v_{1/2})$). We assumed two rotation curves, one with $v_{circ}$=constant (flat overall rotation curve, as in equations (B2) and (B4)), and another one with $v_{circ} = v_{disk}$ as in equation (B2) (baryon dominated disk with a dropping rotation curve, motivated by rotation curve stacks of P. Lang et al. 2016, in preparation). Typically $\log k(x,y)$ is 0.12 dex greater for the flat rotation curve than for the dropping rotation curve. Finally we averaged the results of these two cases, and established the following fitting function

$$\log k(x, y) = -0.082 + 0.091 \times x - (0.06 + 0.244 \times y) \times x^2 - 0.168 \times y \qquad \text{(B11),}$$

which fits all combined data in the interval $x = -0.7$ to $0.7$ and $y = -1.2$ to $-0.15$ to better than ±0.03 dex. For the relevant range in $x$ and $y$, the inferred values of $k(x,y)$ vary from ~1 to ~1.75, where for a thin exponential disk $k(0,-\infty) = 1.19$. These corrections tend to slightly decrease $\lambda \times (j_d/j_{DM})$ for SFGs at the low mass tail, and slightly increase $\lambda \times (j_d/j_{DM})$ for SFGs at the high mass end of our sample. As a default, we omitted these small corrections throughout the paper but discussed where relevant what changes occur if they are applied.

# References

Barnabè, M., Dutton, A. A., Marshall, P. J., et al. 2012, MNRAS, 423, 1073

Barnes, J. & Efstathiou, G. 1987, ApJ, 319, 575

Barro, G., Faber, S. M., Pérez-González, P. G., et al. 2013, ApJ, 765, 104

Barro, G., Faber, S. M., Pérez-González, P. G., et al. 2014a, ApJ, 791, 52

Barro, G., Trump, J. R., Koo, D. C., et al. 2014b, ApJ, 795, 145

Barro, G., Faber, S. M., Dekel, A., et al. 2016, ApJ, 820, 120

Barro, G., Faber, S. M., Koo, D. C., et al. 2015, ApJ, submitted (arXiv:1509.00469)

Behroozi, P. S., Conroy, C., & Wechsler, R. H. 2010, ApJ, 717, 379

Behroozi, P. S., Wechsler, R. H., Wu, H.-Y., et al. 2013a, ApJ, 763, 18

Behroozi, P. S., Marchesini, D., Wechsler, R. H., et al. 2013b, ApJ, 777, 10

Bell, E. F., van der Wel, A., Papovich, C., et al. 2012, ApJ, 753, 167

Bertschinger, E. 1985, ApJS, 58, 39

Béthermin, M., Daddi, E., Magdis, G., et al. 2015, A&A, 573, A113

Bett, P., Eke, V., Frenk, C. S., et al. 2007, MNRAS, 376, 215

Binney, J. & Tremaine, S. 2008, in Galactic Dynamics, eds. J. Binney and S. Tremaine (Princeton, NJ : Princeton Univ. Press), 390

Bonnet, H., Bauvir, B., Wallander, A., et al. 2006, The Messenger, 126, 37

Bournaud, F., Elmegreen, B. G., & Elmegreen, D. M. 2007, ApJ, 670, 237

Bournaud, F., Perret, V., Renaud, F., et al. 2014, ApJ, 780, 57

Bovy, J. & Rix, H.-W. 2013, ApJ, 779, 115

Brammer, G. B., van Dokkum, P. G., Franx, M., et al. 2012, ApJS, 200, 13

Bruce, V. A., Dunlop, J. S., McLure, R. J., et al. 2014a, MNRAS, 444, 1001

Bruce, V. A., Dunlop, J. S., McLure, R. J., et al. 2014b, MNRAS, 444, 1660

Bruzual, G. & Charlot, S. 2003, MNRAS, 344, 1000

Bullock, J. S., Dekel, A., Kolatt, T. S., et al. 2001a, ApJ, 555, 240

Bullock, J. S., Kolatt, T. S., Sigad, Y., et al. 2001b, MNRAS, 321, 559

Burkert, A., Genzel, R., Bouché, N., et al. 2010, ApJ, 725, 2324

Cacciato, M., Dekel, A., & Genel, S. 2012, MNRAS, 421, 818

Calzetti, D, Armus, L., Bohlin, R. C., et al. 2000, ApJ, 533, 682

Cappellari, M., McDermid, R. M., Alatalo, K., et al. 2013, MNRAS 432, 1709

Chabrier, G. 2003, PASP, 115, 763

Cimatti, A., Cassata, P., Pozzetti, L., et al. 2008, A&A, 482, 21

Conroy, C. & Wechsler, R. H. 2009, ApJ, 696, 620

Contini, T., Garilli, B., Le Fèvre, O., et al. 2012, A&A, 539, A91

Courteau, S. & Dutton, A. A. 2015, ApJ, 801, L20

Cowie, L. L., Hu, E. M., & Songaila, A. 1995, AJ, 110, 1576

Cowie, L. L., Hu, E. M., Songaila, A., & Egami, E. 1997, ApJ, 481, L9

Cresci, C., Hicks, E. K. S., Genzel, R., et al. 2009, ApJ, 697, 115

Daddi, E., Alexander, D. M., Dickinson, M., et al. 2007, ApJ, 670, 156

Daddi, E., Bournaud, F., Walter, F., et al. 2010, ApJ, 713, 686

Danovich, M., Dekel, A., Hahn, O., & Teyssier, R. 2012, MNRAS, 422, 1732

Danovich, M., Dekel, A., Hahn, O., Ceverino, D., & Primack, J. 2015, MNRAS 449, 2087

Davies, R. I., Förster Schreiber, N. M., Cresci, G., et al. 2011, ApJ, 741, 69

Dekel, A., Sari, R., & Ceverino, D. 2009, ApJ, 703, 785

Dekel, A. & Burkert, A. 2014, MNRAS, 438, 1870

Dutton, A. A., Conroy, C., van den Bosch, F. C., Prada, F., & More, S. 2010, MNRAS, 407, 2

Dutton, A. A. & van den Bosch, F. C. 2012, MNRAS, 421, 608

Dutton, A. A., Treu, T., Brewer, B. J., et al. 2013, MNRAS, 428, 3183

Eisenhauer, F., Genzel, R., Alexander, T., et al. 2005, ApJ, 628, 246

Elmegreen, B. G. 2009, in "Galaxy Evolution: Emerging Insights and Future Challenges", ASP Conference Series, Vol. 419, p. 23

Elmegreen, B. G., Elmegreen, D. M., Fernandez, M. X., & Lemonias, J. J. 2009, ApJ, 692, 12

Elmegreen, D. M., Elmegreen, B. G., & Hirst, A. C. 2004, ApJ, 604, L21

Engel, H., Tacconi, L. J., Davies, R. I., et al. 2010, ApJ, 724, 233

Épinat, B., Contini, T., Le Fèvre, O., et al. 2009, A&A, 504, 789

Épinat, B., Tasca, L., Amram P., et al. 2012, A&A, 539, A92

Erb, D. K., Shapley, A. E., Pettini, M., et al. 2006, ApJ, 644, 813

Fakhouri, O. & Ma, C.-P. 2008, MNRAS, 386, 577

Fall, S. M. & Efstathiou, G. 1980, MNRAS, 193, 189

Fall, S. M. 1983, IAUS 100, 391

Fall, S. M. & Romanowsky, A. J. 2013, ApJ, 769, L26

Fiacconi, D., Feldmann, R., & Mayer, L. 2015, MNRAS, 446, 1957

Forbes, J. C., Krumholz, M. R., Burkert, A., & Dekel, A. 2014a, MNRAS, 438, 1552

Forbes, J. C., Krumholz, M. R., Burkert, A., & Dekel, A. 2014b, MNRAS, 443, 168

Förster Schreiber, N. M., Genzel, R., Lehnert, M. D., et al. 2006, ApJ, 645, 1062

Förster Schreiber, N. M., Genzel, R., Bouché, N., et al. 2009, ApJ, 706, 1364

Förster Schreiber, N. M., Shapley, A. E., Erb, D. K., et al. 2011a, ApJ, 731, 65

Förster Schreiber, N. M., Shapley, A. E., Genzel, R., et al. 2011b, ApJ,739, 45

Förster Schreiber, N. M., van Dokkum, P. G., Franx, M., et al. 2004, ApJ, 616, 40

Franx, M., van Dokkum, P. G., Förster Schreiber, N. M., et al. 2008, ApJ, 688, 770

Freeman, K. 1970, ApJ, 160, 811

Genel, S., Fall, S. M., Hernquist, L., et al. 2015, ApJ, 804, L40

Genel, S., Genzel, R., Bouché, N., Naab, T., & Sternberg, A. 2009, ApJ, 701, 2002

Genzel, R., Tacconi, L. J., Eisenhauer, F., et al. 2006, Nature, 442, 786

Genzel, R., Burkert, A., Bouché, N., et al. 2008, ApJ, 687, 59

Genzel, R., Newman, S., Jones, T., et al. 2011, ApJ, 733, 101

Genzel, R., Tacconi, L. J., Kurk, J., et al. 2013, ApJ, 773, 68

Genzel, R., Förster Schreiber, N. M., Lang, P., et al. 2014, ApJ, 785, 75

Genzel, R., Tacconi, L. J., Lutz, D., et al. 2015, ApJ, 800, 20

Glazebrook, K. 2013, PASA, 30, 56

Gunn, J. E. & Gott, J. R. III 1972, ApJ, 176, 1

Grogin, N. A., Kocevski, D. D., Faber, S. M., et al. 2011, ApJS, 197, 35

Harrison, C. M., Alexander, D. M., Mullaney, J. R., et al. 2016, MNRAS, 456, 1195

Hetznecker, H. & Burkert, A. 2006, MNRAS, 370, 1905

Hopkins, P. F., Hernquist, L., Cox, T. J., Keres, D., & Wuyts, S. 2009, ApJ, 691, 1424

Hoyle F., 1951, in Burgers J. M., van de Hulst H. C. (eds.), "Problems of Cosmical Aerodynamics", International Union of Theoretical and Applied Mechanics and IAU, p. 195

Immeli, A., Samland, M., Gerhard, O., & Westera, P. 2004b, A&A, 413, 547

Immeli, A., Samland, M., Westera, P., & Gerhard, O. 2004a, ApJ, 611, 20

Jones, T. A., Swinbank, A. M., Ellis, R. S., Richard, J., & Stark, D. P. 2010, MNRAS, 404, 1247

Kassin, S. A., Weiner, B. J., Faber, S. M., et al. 2012, ApJ, 758, 106

Kewley, L. J., Dopita, M. A., Leitherer, C., et al. 2013, ApJ, 774, 100

Kimm, T., Slyz, A., Devriendt, J., & Pichon, C. 2011, MNRAS, 413, 51

Koekemoer, A. M., Faber, S. M., Ferguson, H. C., et al. 2011, ApJS, 197, 36

Komatsu, E., Smith, K. M., Dunkley, J., et al. 2011, ApJS, 192, 18

Krumholz, M. & Burkert, A. 2010, ApJ, 724, 895

Lang, P., Wuyts, S., Somerville, R. S., et al. 2014, ApJ, 788, 11

Larkin, J., Barczys, M., Krabbe, A., et al. 2006, New Astronomy, 50, 362

Law, D. R., Steidel, C. C., Erb, D. K., et al. 2009, ApJ, 697, 2057

Law, D. R., Steidel, C. C., Shapley, A. E., et al. 2012, ApJ, 745, 85

Maccio, A. V., Dutton, A. A., van den Bosch, F. C, et al. 2007, MNRAS, 378, 55

Maccio, A. V., Dutton, A. A., & van den Bosch, F. C. 2008, MNRAS, 391, 1940

Magdis, G. E., Bureau, M., Stott, J. P., et al. 2016, MNRAS, 456, 4533

Mancini, C., Förster Schreiber, N. M., Renzini, A., et al. 2011, ApJ, 743, 86

Maraston, C., Pforr, J., Renzini, A., et al. 2010, MNRAS, 407, 830

Martinsson, T. P. K., Verheijen, M. A. W., Westfall, K. B., et al. 2013a, A&A, 557, A130

Martinsson, T. P. K., Verheijen, M. A. W., Westfall, K. B., et al. 2013b, A&A, 557, A131

Mendel, J. T., Saglia, R. P., Bender, R., et al. 2015, ApJ, 804, L4

Mihos, J. C. & Hernquist, L. 1996, ApJ, 464, 641

Mo, H. J., Mao, S., & White, S. D. M. 1998, MNRAS, 295, 319

Mo, H. J., van den Bosch, F. C., & White, S. D. M., 2010, Galaxy Formation and Evolution, Cambridge University Press, Cambridge

Momcheva, I. G., Brammer, G. B., van Dokkum, P. G., et al. 2015, ApJS, submitted (arXiv:1510.02106)

Moster, B., Somerville, R. S., Maulbetsch, C., et al. 2010, ApJ, 710, 903

Moster, B., Naab, T., & White, S. D. M. 2013, MNRAS, 428, 3121

Navarro, J. F. & Benz, W. 1991, ApJ, 380, 320

Navarro, J. F., Frenk, C. S., & White, S. D. M. 1996, ApJ, 462, 563

Navarro, J. F., Frenk, C. S., & White, S. D. M. 1997, ApJ, 490, 493

Navarro, J. F. & Steinmetz, M. 1997, ApJ, 478, 13

Navarro, J. F. & White, S. D. M. 1994, MNRAS, 267, 401

Nelson, E. J., van Dokkum, P. G., Brammer, G., et al. 2012, ApJ, 747, L28

Nelson, E. J., van Dokkum, P. G., Förster Schreiber, N. M., et al. 2015, ApJ, submitted (arXiv:1507.03999)

Newman, S. F., Genzel, R., Förster Schreiber, N. M., et al. 2013, ApJ, 767, 104

Noguchi, M. 1999, ApJ, 514, 77

Noeske K. G., Weiner B. J., Faber S. M., et al. 2007, ApJ, 660, L43

Noordermeer, E. 2008, MNRAS, 385, 1359

Ostriker, J. P. & Peebles, P. J. E. 1973, ApJ, 186, 467

Ostriker, J. P., Peebles, P. J. E., & Yahil, A. 1974, ApJ, 193, L1

Papovich, C., Dickinson, M., & Ferguson, H. C. 2001, ApJ, 559, 620

Pedrosa, S. E. & Tissera, P. B. 2015, A&A, 584, A43

Peebles, P. J. E. 1969, ApJ, 155, 393

Pichon, C., Pogosyan, D., Kimm, T., et al. 2011, MNRAS, 418, 2493

Rodighiero, G., Cimatti, A., Gruppioni, C., et al. 2010, A&A, 518, L25

Rodighiero, G., Daddi, E., Baronchelli, I., et al. 2011, ApJ, 739, L40

Rodriguez-Gomez, V., Genel, S., Vogelsberger, M., et al. 2015, MNRAS, 449, 49

Romanowsky, A. J. & Fall, S. M. 2012, ApJS, 293, 17

Saintonge, A., Lutz, D., Genzel, R., et al. 2013, ApJ, 778, 2

Sanders, R. L., Shapley, A. E., Kriek, M., et al. 2015, ApJ, 799, 138

Sargent, M. Y., Daddi, E., Béthermin, M., et al. 2014, ApJ, 793, 19

Schiminovich, D., Wyder, T. K., Martin, D. C., et al. 2007, ApJS, 173, 315

Shapley, A. E., Steidel, C. C., Erb, D. K., et al. 2005, ApJ, 626, 698

Sharples, R. M., Ramsay, S. K., Davies, R., & Lehnert, M. 2008, in „The 2007 ESO Instrument Calibration Workshop", ESO Astrophysics Symposia (Springer, Heidelberg), 311

Sharples, R., Bender, R., Agudo, A., et al. 2012, in „Ground-based and Airborne Instrumentation for Astronomy IV", proceedings of the SPIE 8446, 9

Skelton, R. E., Whitaker, K. E., Momcheva, I. G., et al. 2014, ApJS, 214, 24

Sobral, D., Swinbank, A. M., Stott, J. P., et al. 2013, ApJ, 779, 139

Sofue, Y. & Rubin, V. 2001, ARAA, 39, 137

Speagle, J. S., Steinhardt, C. L., Capak, P. L., & Silverman, J. D. 2014, ApJS, 214, 15

Steidel, C. C., Rudie, G. C., Strom, A. L., et al. 2014, ApJ, 795, 165

Stewart, K. R., Brooks, A. M., Bullock, J. S., et al. 2013, ApJ, 769, 74

Stewart, K. R., Kaufmann, T., Bullock, J. S., et al. 2011, ApJ, 738, 39

Stott, J. P., Sobral, D., Bower, R., et al. 2013, MNRAS, 436, 1130

Stott, J. P., Sobral, D., Swinbank, A. M., et al. 2014, MNRAS, 443, 2695

Stott, J. P., Swinbank, A. M., Johnson, H. L., et al. 2016, MNRAS, 457, 1888

Swinbank, A. M., Sobral, D., Smail, I., et al. 2012, MNRAS, 426, 935

Tacchella, S., Lang, P., Carollo, C. M., et al. 2015a, ApJ, 802, 101

Tacchella, S., Carollo, C. M., Renzini, A., et al. 2015b, Science, 348, 314

Tacchella, S., Dekel, A., Carollo, C. M., et al. 2016, MNRAS, 458, 242

Tacconi, L. J., Genzel, R., Smail, I., et al. 2008 ApJ, 680, 246

Tacconi, L. J., Genzel, R., Neri, R., et al. 2010, Nature, 463, 781

Tacconi, L. J., Neri, R., Genzel, R., et al. 2013, ApJ, 768, 74

Teklu, A. F., Remus, R.-S., Dolag, K., et al. 2015, ApJ, 812, 29

Übler, H., Naab, T., Oser, L., et al. 2014, MNRAS, 443, 2092

van den Bergh, S., Abraham, R. G., Ellis, R. S., et al. 1996, AJ, 112, 359

van der Kruit, P. C. & Allen, R. J. 1978, ARAA, 16, 103

van der Wel, A., Franx, M., van Dokkum, P. G., et al. 2014a, ApJ, 788, 28

van der Wel, A. Chang, Y.-Y., Bell, E. F., et al. 2014b, ApJ, 792, L6

van Dokkum, P. G., Nelson, E. J., Franx, M., et al. 2015, ApJ, 813, 23

van Starkenburg, L., van der Werf, P. P., Franx, M., et al. 2008, A&A, 488, 99

Wellons, S., Torrey, P., Ma, C.-P., et al. 2015, MNRAS, 449, 361

Whitaker, K. E., van Dokkum, P. G., Brammer, G., & Franx, M. 2012, ApJ, 754, L29

Whitaker, K. E., Franx, M., Leja, J., et al. 2014, ApJ, 795, 104

Whitaker, K. E., Franx, M., Bezanson, R., et al. 2015, ApJ, 811, L12

White, S. D. M. 1984, ApJ, 286, 38

Wisnioski, E., Glazebrook, K., Blake, C., et al. 2011, MNRAS, 417, 2601

Wisnioski, E., Glazebrook, K., Blake, C., et al. 2012, MNRAS, 422, 339

Wisnioski, E., Förster Schreiber, N. M., Wuyts, S., et al. 2015, ApJ, 799, 209

Wizinowich, P. L., Le Mignant, D., Bouchez, A. H., et al. 2006, PASP, 118, 297

Wright, S. A., Larkin, J. E., Barczys, M., et al. 2007, ApJ, 658, 78

Wright, S. A., Larkin, J. E., Law, D. R., et al. 2009, ApJ, 699, 421

Wright, S. A., Ma, C., Larkin, J., & Graham, J. 2011, Bulletin of the American Astronomical Society, 43, 2011

Wuyts, E., Kurk, J., Förster Schreiber, N. M., et al. 2014, ApJ, 789, L40

Wuyts, E., Wisnioski, E., Fossati, M., et al. 2016, ApJ, submitted (arXiv:1603:01139)

Wuyts, S., Labbé, I., Franx, M., et al. 2007, ApJ, 655, 51

Wuyts, S., Förster Schreiber, N. M., Lutz, D., et al. 2011a, ApJ, 738, 106

Wuyts, S., Förster Schreiber, N. M., van der Wel, A., et al. 2011b, ApJ, 742, 96

Wuyts, S., Förster Schreiber, N. M., Genzel, R., et al. 2012, ApJ, 753, 114

Wuyts, S., Förster Schreiber, N. M., Genzel, R., et al. 2013, ApJ, 779, 135

Wuyts, S., Förster Schreiber, N. M., Wisnioski, E., et al. 2016, ApJ, submitted (arXiv:1603.03432)

Zahid, H. J., Kashino, D., Silverman, J. D., et al. 2014, ApJ, 792, 75

Zahid, H. J., Kewley, L. J., & Bresolin, F. 2011, ApJ, 730, 137

Zavala, J., Frenk, C. S., Bower, R., et al. 2015, MNRAS, submitted (arXiv:1512.02636)

Zolotov, A., Dekel, A., Mandelker, N., et al. 2015, MNRAS, 450, 2327